\documentclass{aa}  

\usepackage{graphicx}
\usepackage{txfonts}
\usepackage{color}
\usepackage{multirow}

\newcommand{\tablenotea}[1]{\parbox{16.9cm}{\indent \footnotesize{#1}}}
\newcommand{\tablenoteb}[1]{\parbox{18.1cm}{\indent \footnotesize{#1}}}
\newcommand{\tablenotec}[1]{\parbox{18.5cm}{\indent \footnotesize{#1}}}
\newcommand{\tablenoted}[1]{\parbox{13.3cm}{\indent \footnotesize{#1}}}
\newcommand{\cpl}{Chem. Phys. Lett.}
\newcommand{\chemrev}{Chem. Rev.}

\newcommand{\irpc}{Int. Rev. Phys. Chem.}

\newcommand{\jpb}{J. Phys. B: At. Mol. Opt. Phys.}

\newcommand{\jpca}{J. Phys. Chem. A}

\newcommand{\nature}{Nature}

\newcommand{\pss}{Planet. Space Sci.}
\newcommand{\rscadv}{RSC Adv.}

\begin{document}

\title{The abundance and excitation of molecular anions\\ in interstellar clouds\thanks{Based on observations carried out with the Yebes 40m telescope (projects 19A003, 20A014, 20A016, 20B010, 20D023, 21A006, 21A011, 21D005, 22B023, and 23A024) and the IRAM 30m telescope. The 40m radio telescope at Yebes Observatory is operated by the Spanish Geographic Institute (IGN; Ministerio de Transportes, Movilidad y Agenda Urbana). IRAM is supported by INSU/CNRS (France), MPG (Germany), and IGN (Spain).}}

\titlerunning{Molecular anions in the ISM}
\authorrunning{Ag\'undez et al.}

\author{M.~Ag\'undez\inst{1}, N.~Marcelino\inst{2,3}, B.~Tercero\inst{2,3}, I. Jim\'enez-Serra\inst{4}, \and J.~Cernicharo\inst{1}}

\institute{
Instituto de F\'isica Fundamental, CSIC, Calle Serrano 123, E-28006 Madrid, Spain\\ \email{marcelino.agundez@csic.es} 
\and
Observatorio Astron\'omico Nacional, IGN, Calle Alfonso XII 3, E-28014 Madrid, Spain 
\and
Observatorio de Yebes, IGN, Cerro de la Palera s/n, E-19141 Yebes, Guadalajara, Spain
\and
Centro de Astrobiolog\'ia (CSIC/INTA), Ctra. de Torrej\'on a Ajalvir km 4, 28806, Torrej\'on de Ardoz, Spain
}

\date{Received; accepted}

 
\abstract
{We report new observations of molecular anions with the Yebes\,40m and IRAM\,30m telescopes toward the cold dense clouds TMC-1\,CP, Lupus-1A, L1527, L483, L1495B, and L1544. We detected for the first time C$_3$N$^-$ and C$_5$N$^-$ in Lupus-1A and C$_4$H$^-$ and C$_6$H$^-$ in L483. In addition, we report new lines of C$_6$H$^-$ toward the six targeted sources, of C$_4$H$^-$ toward TMC-1\,CP, Lupus-1A, and L1527, and of C$_8$H$^-$ and C$_3$N$^-$ in TMC-1\,CP. Excitation calculations using recently computed collision rate coefficients indicate that the lines of anions accessible to radiotelescopes run from subthermally excited to thermalized as the size of the anion increases, with the degree of departure from thermalization depending on the H$_2$ volume density and the line frequency. We noticed that the collision rate coefficients available for the radical C$_6$H cannot explain various observational facts, which advises for a revisitation of the collision data for this species. The observations presented here, together with observational data from the literature, are used to model the excitation of interstellar anions and to constrain their abundances. In general, the anion-to-neutral ratios derived here agree within 50\,\% (a factor of two at most) with literature values, when available, except for the C$_4$H$^-$/C$_4$H ratio, which shows higher differences due to a revision of the dipole moment of C$_4$H. From the set of anion-to-neutral abundance ratios derived two conclusions can be drawn. First, the C$_6$H$^-$/C$_6$H ratio shows a tentative trend in which it increases with increasing H$_2$ density, as expected from theoretical grounds. And second, it is incontestable that the higher the molecular size the higher the anion-to-neutral ratio, which supports a formation mechanism based on radiative electron attachment. Nonetheless, calculated rate coefficients for electron attachment to the medium size species C$_4$H and C$_3$N are probably too high and too low, respectively, by more than one order of magnitude. 
}


\maketitle

\section{Introduction}

The discovery of negatively charged molecular ions in space has been a relatively recent finding \citep{McCarthy2006}. To date the inventory of molecular anions detected in interstellar and circumstellar clouds consists of four hydrocarbon anions, C$_4$H$^-$ \citep{Cernicharo2007}, C$_6$H$^-$ \citep{McCarthy2006}, C$_8$H$^-$ \citep{Brunken2007a,Remijan2007}, and C$_{10}$H$^-$ \citep{Remijan2023}, and four nitrile anions, CN$^-$ \citep{Agundez2010}, C$_3$N$^-$ \citep{Thaddeus2008}, C$_5$N$^-$ \citep{Cernicharo2008}, and C$_7$N$^-$ \citep{Cernicharo2023a}. The astronomical detection of most of these species has been possible thanks to the laboratory characterization of their rotational spectrum \citep{McCarthy2006,Gupta2007,Gottlieb2007,Thaddeus2008}. However, the astronomical detection of C$_5$N$^-$, C$_7$N$^-$, and C$_{10}$H$^-$ is based on high level \textit{ab initio} calculations and astrochemical arguments \citep{Botschwina2008,Cernicharo2008,Cernicharo2020,Cernicharo2023a,Remijan2023}. In fact, in the case of C$_{10}$H$^-$ it is not yet clear whether the identified species is C$_{10}$H$^-$ or C$_9$N$^-$ \citep{Pardo2023}.

The current situation is such that there is only one astronomical source where the eight molecular anions have been observed, the carbon-rich circumstellar envelope IRC\,+10216 \citep{McCarthy2006,Cernicharo2007,Cernicharo2008,Remijan2007,Thaddeus2008,Agundez2010,Cernicharo2023a,Pardo2023}, while the first negative ion discovered, C$_6$H$^-$ \citep{McCarthy2006}, continues to be the most widely observed in astronomical sources \citep{Sakai2007,Sakai2010,Gupta2009,Cordiner2011,Cordiner2013}.

Observations indicate that along each of the series C$_{2n+2}$H$^-$ and C$_{2n-1}$N$^-$ (with $n$\,=\,1, 2, 3, 4) the anion-to-neutral abundance ratio increases with increasing molecular size \citep{Millar2017}. This is expected according to the formation mechanism originally proposed by \cite{Herbst1981}, which involves the radiative electron attachment to the neutral counterpart of the anion \citep{Herbst2008,Carelli2013}. However, the efficiency of this mechanism in interstellar space has been disputed \citep{Khamesian2016}, and alternative formation mechanisms have been proposed \citep{Gianturco2016}. Currently there is not yet consensus on the formation mechanism of molecular anions in space (see discussion in \citealt{Millar2017}). Moreover, detections of negative ions other than C$_6$H$^-$ in interstellar clouds are scarce, and thus our view of the abundance of the different anions in interstellar space is statistically very limited.

Apart from the anion-to-anion behavior it is also interesting to know which is the source-to-source behavior. That is, how does the abundance of anions behave from one source to another. Based on C$_6$H$^-$ detections, the C$_6$H$^-$/C$_6$H abundance ratio seems to increase with increasing H$_2$ volume density \citep{Sakai2007,Agundez2008,Cordiner2013}, which is expected from chemical considerations (e.g., \citealt{Flower2007}; see also Sect.~\ref{sec:discussion}). However, most anion detections in interstellar clouds have been based on one or two lines and their abundances have been estimated assuming that their rotational levels are populated according to local thermodynamic equilibrium (LTE), which may not be a good assumption given the large dipole moments, and thus high critical densities, of anions. Recently, rate coefficients for inelastic collisions with H$_2$ or He have been calculated for C$_2$H$^-$ \citep{Dumouchel2012,Gianturco2019,Franz2020,Toumi2021,Dumouchel2023}, C$_4$H$^-$ \citep{Senent2019,Balanca2021}, C$_6$H$^-$ \citep{Walker2016,Walker2017}, CN$^-$ \citep{Klos2011,Gonzalez-Sanchez2020}, C$_3$N$^-$ \citep{Lara-Moreno2017,Lara-Moreno2019,Tchakoua2018}, and C$_5$N$^-$ \citep{Biswas2023}, which makes it possible to study the excitation of anions in the interstellar medium.

Here we report new detections of anions in interstellar sources. Concretely, we detected C$_3$N$^-$ and C$_5$N$^-$ in Lupus-1A and C$_6$H$^-$ and C$_4$H$^-$ in L483. We also present the detection of new lines of C$_4$H$^-$, C$_6$H$^-$, C$_8$H$^-$, C$_3$N$^-$, and C$_5$N$^-$ in interstellar clouds where these anions have been already observed. We use the large observational dataset from this study, together with that available from the literature, to review the observational status of anions in interstellar clouds and to carry out a comprehensive analysis of the abundance and excitation of anions in the interstellar medium.

\section{Observations}

\subsection{Yebes\,40m and IRAM\,30m observations from this study} \label{sec:observations}

The observations of cold dark clouds presented in this study were carried out with the Yebes\,40m and IRAM\,30m telescopes. We targeted the starless core TMC-1 at the cyanopolyyne peak position (hereafter TMC-1\,CP)\footnote{TMC-1\,CP: $\alpha_{J2000}=4^{\rm h} 41^{\rm  m} 41.9^{\rm s}$ and $\delta_{J2000}=+25^\circ 41' 27.0''$}, the starless core Lupus-1A\footnote{Lupus-1A: $\alpha_{J2000}=15^{\rm h} 42^{\rm  m} 52.4^{\rm s}$ and $\delta_{J2000}=$-$34^\circ 07' 53.5''$}, the prestellar cores L1495B\footnote{L1495B: $\alpha_{J2000}=4^{\rm h} 15^{\rm  m} 41.8^{\rm s}$ and $\delta_{J2000}=+28^\circ 47' 46.0''$} and L1544\footnote{L1544: $\alpha_{J2000}=5^{\rm h} 4^{\rm  m} 18.0^{\rm s}$ and $\delta_{J2000}=+25^\circ 11' 10.0''$}, and the dense cores L1527\footnote{L1527: $\alpha_{J2000}=4^{\rm h} 39^{\rm  m} 53.9^{\rm s}$ and $\delta_{J2000}=+26^\circ 03' 11.0''$} and L483\footnote{L483: $\alpha_{J2000}=18^{\rm h} 17^{\rm  m} 29.8^{\rm s}$ and $\delta_{J2000}=$-$4^\circ 39' 38.3''$}, which host a Class\,0 protostar. All observations were done using the frequency switching technique to maximize the on-source telescope time and to improve the sensitivity of the spectra.

The Yebes\,40m observations consisted in a full scan of the $Q$ band (31-50 GHz) acquired in a single spectral setup with a 7\,mm receiver, which was connected to a fast Fourier transform spectrometer that provides a spectral resolution of 38 kHz \citep{Tercero2021}. The data of TMC-1\,CP are part of the on-going QUIJOTE line survey \citep{Cernicharo2021}. The spectra used here were obtained between November 2019 and November 2022 and contain a total of 758 h of on-source telescope time in each polarization (twice this value after averaging both polarizations). Two frequency throws of 8 and 10 MHz were used. The sensitivity ranges from 0.13 to 0.4 mK in antenna temperature. The data of L1544 were taken between October and December 2020 toward the position of the methanol peak of this core, where complex organic molecules have been detected \citep{Jimenez-Serra2016}, and are part of a high-sensitivity Q-band survey (31 h on-source; Jim\'enez-Serra et al. in prep.). The data for the other sources were obtained from July 2020 to February 2023 for L483 (the total on-source telescope time is 103 h), from May to November 2021 for L1527 (40 h on-source), from July 2021 to January 2023 for Lupus-1A (120 h on-source), and from September to November 2021 for L1495B (45 h on-source). Different frequency throws were adopted depending on the observing period, which resulted from tests done at the Yebes\,40m telescope to find the optimal frequency throw. We used frequency throws of 10 MHz and 10.52 MHz for L483, 10 MHz for L1544, 8 MHz for L1527, and 10.52 MHz for Lupus-1A and L1495B. The antenna temperature noise levels, after averaging horizontal and vertical polarizations, are in the range 0.4-1.0 mK for L483, 1.3-1.8 mK for L1544, 0.7-2.7 mK for L1527, 0.7-2.8 mK for Lupus-1A, and 0.8-2.6 mK for L1495B.

The observations carried out with the IRAM\,30m telescope used the 3\,mm EMIR receiver connected to a fast Fourier transform spectrometer that provides a spectral resolution of 49 kHz. Different spectral regions within the 3\,mm band (72-116 GHz) were covered depending on the source. The data of TMC-1\,CP consist of a 3\,mm line survey \citep{Marcelino2007,Cernicharo2012} and spectra observed in 2021 \citep{Agundez2022,Cabezas2022}. The data of L483 consists of a line survey in the 80-116 GHz region (see \citealt{Agundez2019}), together with data in the 72-80 GHz region, which are described in \cite{Cabezas2021}. Data of Lupus-1A, L1495B, L1521F, L1251A, L1512, L1172, and L1389 were observed from September to November 2014 during a previous search for molecular anions at mm wavelengths (see \citealt{Agundez2015}). Additional data of Lupus-1A were gathered during 2021 and 2022 during a project aimed to observe H$_2$NC \citep{Agundez2023}. In the case of L1527, the IRAM\,30m data used were observed in July and August 2007 with the old ABCD receivers connected to an autocorrelator that provided spectral resolutions of 40 or 80 kHz \citep{Agundez2008}.

The half power beam width (HPBW) of the Yebes\,40m telescope is in the range 35-57\,$''$ in the $Q$ band, while that of the IRAM\,30m telescope ranges between 21\,$''$ and 34\,$''$ in the 3\,mm band. The beam size can be fitted as a function of frequency as HPBW($''$)\,=\,1763/$\nu$(GHz) for the Yebes\,40m telescope and as HPBW($''$)\,=\,2460/$\nu$(GHz) for the IRAM\,30m telescope. Therefore, the beam size of the IRAM\,30m telescope at 72 GHz is similar to that of the Yebes\,40m at 50 GHz. The intensity scale in both the Yebes\,40m and IRAM\,30m telescopes is antenna temperature, $T_A^*$, for which we estimate a calibration error of 10\,\%. To convert antenna temperature into main beam brightness temperature see foot of Table~\ref{table:lines_anion}. All data were analyzed using the program CLASS of the GILDAS software\footnote{https://www.iram.fr/IRAMFR/GILDAS/}.

\begin{figure*}
\centering
\includegraphics[angle=0,width=\textwidth]{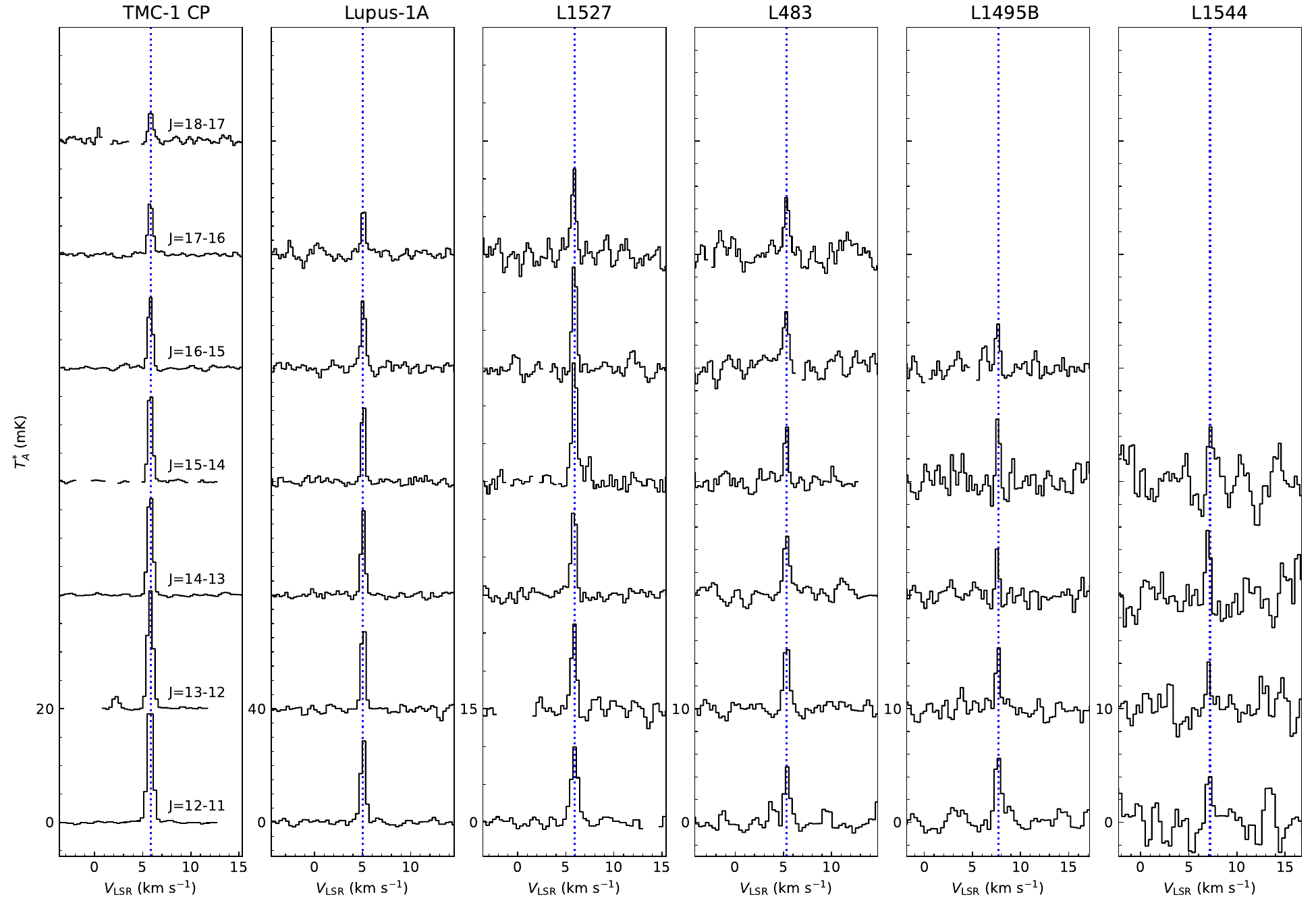}
\caption{Lines of C$_6$H$^-$ observed in this work toward six cold dense clouds using the Yebes\,40m telescope. See line parameters in Table~\ref{table:lines_anion}.}
\label{fig:lines_c6hm}
\end{figure*}

\begin{figure*}
\centering
\includegraphics[angle=0,width=0.80\textwidth]{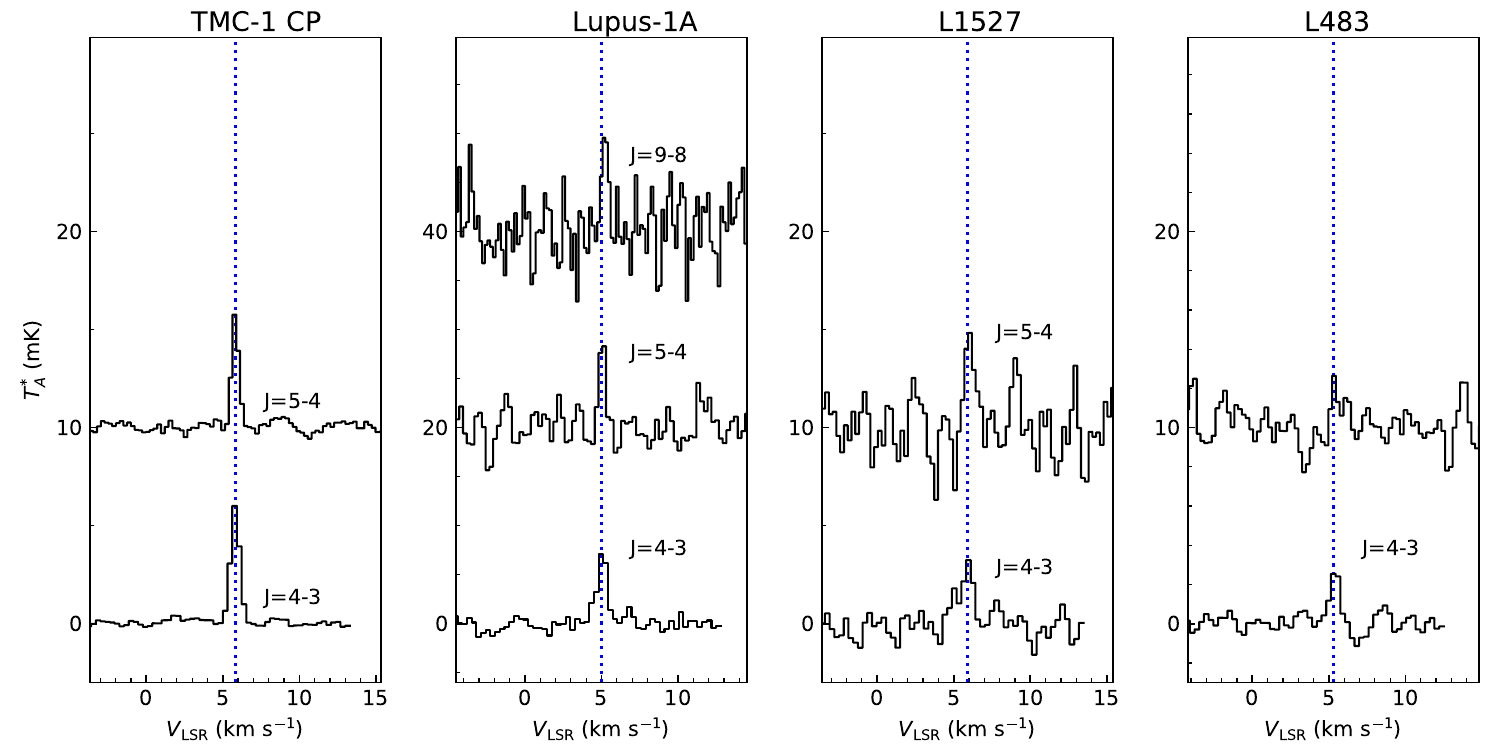}
\caption{Lines of C$_4$H$^-$ observed in this work toward TMC-1\,CP, Lupus-1A, L1527, and L483 using the Yebes\,40m and IRAM\,30m telescopes. See line parameters in Table~\ref{table:lines_anion}.}
\label{fig:lines_c4hm}
\end{figure*}

\begin{figure*}
\centering
\includegraphics[angle=0,width=0.95\textwidth]{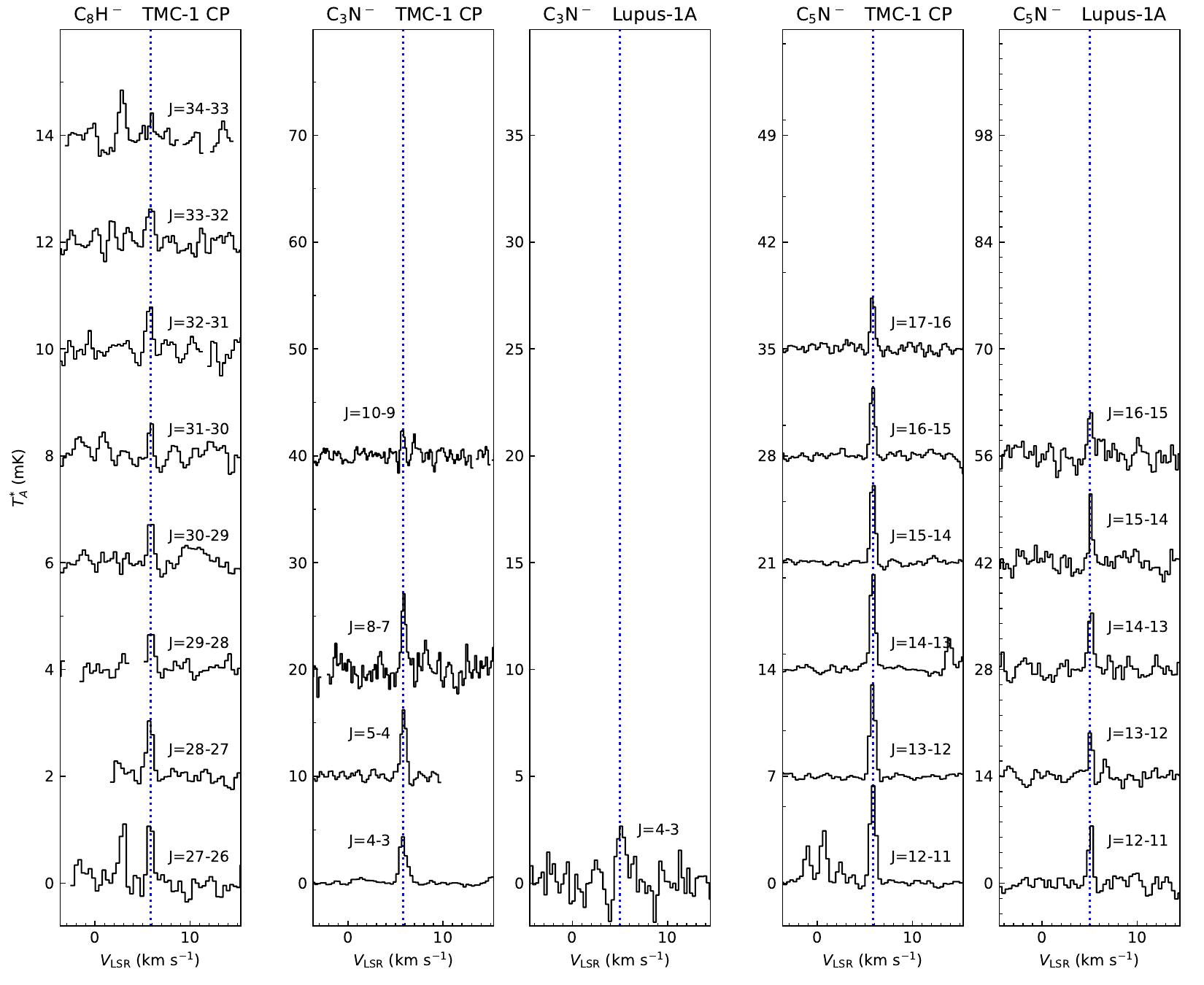}
\caption{Lines of C$_8$H$^-$ observed toward TMC-1\,CP and lines of the nitrile anions C$_3$N$^-$ and C$_5$N$^-$ observed toward TMC-1\,CP and Lupus-1A using the Yebes\,40m and IRAM\,30m telescopes. See line parameters in Table~\ref{table:lines_anion}.}
\label{fig:lines_rest}
\end{figure*}

\subsection{Observational dataset of anions in dark clouds} \label{sec:observational_dataset}

In Table~\ref{table:lines_anion} we compile the line parameters of all the lines of negative molecular ions detected toward cold dark clouds, including lines from this study and from the literature. The line parameters of C$_7$N$^-$ observed toward TMC-1\,CP are given in \cite{Cernicharo2023a} and are not repeated here. In the case of C$_{10}$H$^-$ in TMC-1\,CP we do not include line parameters here because the detection by \cite{Remijan2023} is not based on individual lines but on spectral stack of many lines. The lines of molecular anions presented in this study are shown in Fig.~\ref{fig:lines_c6hm} for C$_6$H$^-$, Fig.~\ref{fig:lines_c4hm} for C$_4$H$^-$, and Fig.~\ref{fig:lines_rest} for the remaining anions, i.e., C$_8$H$^-$, C$_3$N$^-$, and C$_5$N$^-$. Since we are interested in the determination of anion-to-neutral abundance ratios, we also need the lines of the corresponding neutral counterpart of each molecular anion, which are the radicals C$_4$H, C$_6$H, C$_8$H, C$_3$N, and C$_5$N. The velocity-integrated intensities of the lines of these species are given in Table~\ref{table:lines_neutral}.

According to the literature, the most prevalent molecular anion, C$_6$H$^-$, has been detected in 11 cold dark clouds: TMC-1\,CP \citep{McCarthy2006}, L1527 and Lupus-1A \citep{Sakai2007,Sakai2010}, L1544 and L1521F \citep{Gupta2009}, and L1495B, L1251A, L1512, L1172, L1389, and TMC-1\,C \citep{Cordiner2011,Cordiner2013}. All these detections were based on two individual or stacked lines lying in the frequency range 11-31 GHz (see Table~\ref{table:lines_anion}). Here we present additional lines of C$_6$H$^-$ in the $Q$ band for TMC-1\,CP, Lupus-1A, L1527, L1495B, and L1544, together with the detection of C$_6$H$^-$ in a new source, L483, through six lines lying in the $Q$ band (see Fig.~\ref{fig:lines_c6hm}).

Molecular anions different to C$_6$H$^-$ have turned out to be more difficult to detect as they have been only seen in a few sources. For example, C$_4$H$^-$ has been only detected in three dark clouds, L1527 \citep{Agundez2008}, Lupus-1A \citep{Sakai2010}, and TMC-1\,CP \citep{Cordiner2013}. These detections rely on one or two lines (see Table~\ref{table:lines_anion}). Here we report the detection of two additional lines of C$_4$H$^-$ in the $Q$ band toward these three sources, together with the detection of C$_4$H$^-$ in one new source, L483 (see Fig.~\ref{fig:lines_c4hm}).

The hydrocarbon anion C$_8$H$^-$ has been observed in two interstellar sources. \cite{Brunken2007a} reported the detection of four lines in the 12-19 GHz frequency range toward TMC-1\,CP, while \cite{Sakai2010} reported the detection of this anion in Lupus-1A through two stacked lines at 18.7 and 21.0 GHz (see Table~\ref{table:lines_anion}). Thanks to our Yebes\,40m data, we present new lines of C$_8$H$^-$ in the $Q$ band toward TMC-1\,CP (see Fig.~\ref{fig:lines_rest}).

Finally, the nitrile anions C$_3$N$^-$ and C$_5$N$^-$ have resulted to be quite elusive as they have been only seen in one cold dark cloud, TMC-1\,CP \citep{Cernicharo2020}. Here we present the same lines of C$_3$N$^-$ and C$_5$N$^-$ reported in \cite{Cernicharo2020} in the $Q$ band, but with improved signal-to-noise ratios, plus two additional lines of C$_3$N$^-$ in the 3\,mm band. We also present the detection of C$_3$N$^-$ and C$_5$N$^-$ in one additional source, Lupus-1A (see Fig.~\ref{fig:lines_rest}).

\section{Physical parameters of the sources} \label{sec:sources}

\begin{table*}
\small
\caption{Source parameters.}
\label{table:sources}
\centering
\begin{tabular}{lcllccccllccll}
\hline \hline
\multicolumn{1}{l}{Source} & & \multicolumn{2}{c}{Type} & & \multicolumn{1}{c}{$\Delta v$} & & \multicolumn{3}{c}{$T_k$} & & \multicolumn{3}{c}{$n$(H$_2$)} \\
\cline{3-4} \cline{6-6} \cline{8-10} \cline{12-14}
 & & \multicolumn{1}{l}{} & \multicolumn{1}{l}{Ref} & & \multicolumn{1}{c}{(km s$^{-1}$)} & & \multicolumn{1}{c}{(K)} & \multicolumn{1}{l}{Method} & \multicolumn{1}{l}{Ref} & & \multicolumn{1}{c}{(cm$^{-3}$)} & \multicolumn{1}{l}{Method} & \multicolumn{1}{l}{Ref} \\
\hline
TMC-1\,CP & & starless   & (1) & & 0.60 & &  9 & CH$_3$CCH, CH$_3$C$_4$H   & (8)       & & 1.0\,$\times$\,10$^4$    & HC$_3$N with $^{13}$C & (8)  \\
Lupus-1A  & & starless   & (2) & & 0.50 & & 11 & CH$_3$CCH                 & (8)       & & 1.8\,$\times$\,10$^4$    & HC$_3$N with $^{13}$C & (8)  \\
L1527     & & protostar  & (3) & & 0.60 & & 14 & CH$_3$CCH                 & (3,8)     & & $>$\,1\,$\times$\,10$^5$ & HC$_3$N with $^{13}$C & (8)  \\
L483      & & protostar  & (4) & & 0.52 & & 12 & CH$_3$CCH                 & (8)       & & 5.6\,$\times$\,10$^4$    & HC$_3$N with $^{13}$C & (8)  \\
L1495B    & & prestellar & (5) & & 0.50 & &  9 & CH$_3$CCH                 & (8)       & & 1.6\,$\times$\,10$^4$    & HCC$^{13}$CN          & (8)  \\
L1544     & & prestellar & (5) & & 0.60 & & 10 & NH$_3$, C$^{17}$O, SO$_2$ & (9,10,11) & & 2\,$\times$\,10$^4$      & SO, SO$_2$            & (11,12) \\
L1521F    & & prestellar & (5) & & 0.45 & &  9 & CH$_3$CCH, NH$_3$         & (8,13)    & & 1\,$\times$\,10$^4$    & HCCNC                 & (8)  \\
L1251A    & & protostar  & (6) & & 0.40 & & 10 & HC$_3$N hfs               & (6)       & & 2.1\,$\times$\,10$^4$    & HC$_3$N               & (6)  \\
L1512     & & starless   & (5) & & 0.30 & & 10 & HC$_3$N hfs               & (6)       & & 2.6\,$\times$\,10$^4$    & HC$_3$N               & (6)  \\
L1172     & & protostar  & (7) & & 0.55 & & 10 & HC$_3$N hfs               & (6)       & & 7.5\,$\times$\,10$^4$    & HC$_3$N               & (6)  \\
L1389     & & protostar  & (6) & & 0.40 & & 10 & HC$_3$N hfs               & (6)       & & 5.2\,$\times$\,10$^4$    & HC$_3$N               & (6)  \\
TMC-1\,C  & & starless   & (5) & & 0.18 & & 10 & HC$_3$N hfs               & (6)       & & 1.1\,$\times$\,10$^4$    & HC$_3$N               & (6)  \\
\hline
\end{tabular}
\tablenoteb{\\
References: (1) \cite{Suzuki1992}. (2) \cite{Sakai2010}. (3) \cite{Sakai2008}. (4) \cite{Agundez2019}. (5) \cite{Crapsi2005}. (6) \cite{Cordiner2013}. (7) \cite{Visser2002}. (8) This work (see text). (9) \cite{Tafalla2002}. (10) \cite{Bacmann2002}. (11) \cite{Vastel2018}. (12) \cite{Punanova2018}. (13) \cite{Codella1997}.
}
\end{table*}

The interstellar clouds where molecular anions have been detected are 12 in total and comprise cold dense cores in different evolutionary stages, such as starless, prestellar, and protostellar (see Table~\ref{table:sources}). The classification as protostellar cores is evident in the cases of L1527 and L483 as the targeted positions are those of the infrared sources IRAS\,04368+2557 and IRAS\,18148$-$0440, respectively \citep{Sakai2008,Agundez2019}. We also classified L1251A, L1172, and L1389 as protostellar sources based on the proximity of an infrared source (L1251A\,IRS3, CB17\,MMS, and IRAS\,21017+6742, respectively) to the positions targeted by \cite{Cordiner2013}. The differentiation between starless and prestellar core is in some cases more ambiguous. In those cases we followed the criterion based on the N$_2$D$^+$/N$_2$H$^+$ column density ratio by \cite{Crapsi2005}. In any case, for our purposes it is not very important whether a given core is starless or prestellar.

To study the abundance and excitation of molecular anions in these 12 interstellar sources through non-LTE calculations we need to know which are the physical parameters of the clouds, mainly the gas kinetic temperature and the H$_2$ volume density, but also the emission size of anions and the linewidth. The adopted parameters are summarized in Table~\ref{table:sources}.

Given that C$_6$H$^-$ has not been mapped in any interstellar cloud to date, it is not known whether the emission of molecular anions in each of the 12 sources is extended compared to the telescope beam sizes, which are in the range 21-67\,$''$ for the Yebes\,40m, IRAM\,30m, and GBT telescopes at the frequencies targeted for the observations of anions. Therefore one has to rely on maps of related species. In the case of TMC-1\,CP we assume that anions are distributed in the sky as a circle with a diameter of 80\,$''$ based on the emission distribution of C$_6$H mapped by \cite{Fosse2001}. Recent maps carried out with the Yebes\,40m telescope \citep{Cernicharo2023b} support the previous results of \cite{Fosse2001}. For the remaining 11 sources, the emission distribution of C$_6$H is not known and thus we assume that the emission of anions is extended with respect to the telescope beam. This assumption is supported by the extended nature of HC$_3$N emission in the cases of L1495B, L1251A, L1512, L1172, L1389, and TMC-1\,C, according to the maps presented by \cite{Cordiner2013}, and of multiple molecular species, including C$_4$H, in L1544, according to the maps reported by \cite{Spezzano2017}.

The linewidth adopted for each source (see Table~\ref{table:sources}) was calculated as the arithmetic mean of the values derived for the lines of C$_6$H$^-$ in the $Q$ band for TMC-1\,CP, Lupus-1A, L1527, L1495B, and L1544. In the case of L483 we adopted the value derived by \cite{Agundez2019} from the analysis of all the lines in the 3\,mm band. For L1521F, L1251A, L1512, L1172, and L1389 the adopted linewidths come from IRAM\,30m observations of CH$_3$CCH in the 3\,mm band (see Sect.~\ref{sec:observations}). Finally, for TMC-1\,C we adopted as linewidth that derived for HC$_3$N by \cite{Cordiner2013}.

The gas kinetic temperature was determined for some of the sources from the $J$\,=\,5-4 and $J$\,=\,6-5 rotational transitions of CH$_3$CCH, which lie around 85.4 and 102.5 GHz, respectively. We have IRAM\,30m data of these lines for TMC-1\,CP, Lupus-1A, L483, L1495B, and L1521F, while for L1527 we used the data obtained with the Nobeyama\,45m telescope by \cite{Yoshida2019}. Typically, the $K$\,=\,0, 1, and 2 components are detected, which allow us to use the line intensity ratio between the $K$\,=\,1 and $K$\,=\,2 components, belonging to the $E$ symmetry species, to derive the gas kinetic temperature. Since transitions with $\Delta K$\,$\ne$\,0 are radiatively forbidden, the relative populations of the $K$\,=\,1 and $K$\,=\,2 levels are controlled by collisions with H$_2$ and thus are thermalized at the kinetic temperature of H$_2$. We do not use the $K$\,=\,0 component because it belongs to a different symmetry species, $A$, and interconversion between $A$ and $E$ species is expected to be slow in cold dense clouds and thus their relative populations may not necessarily reflect the gas kinetic temperature.

\begin{figure*}
\centering
\includegraphics[angle=0,width=0.325\textwidth]{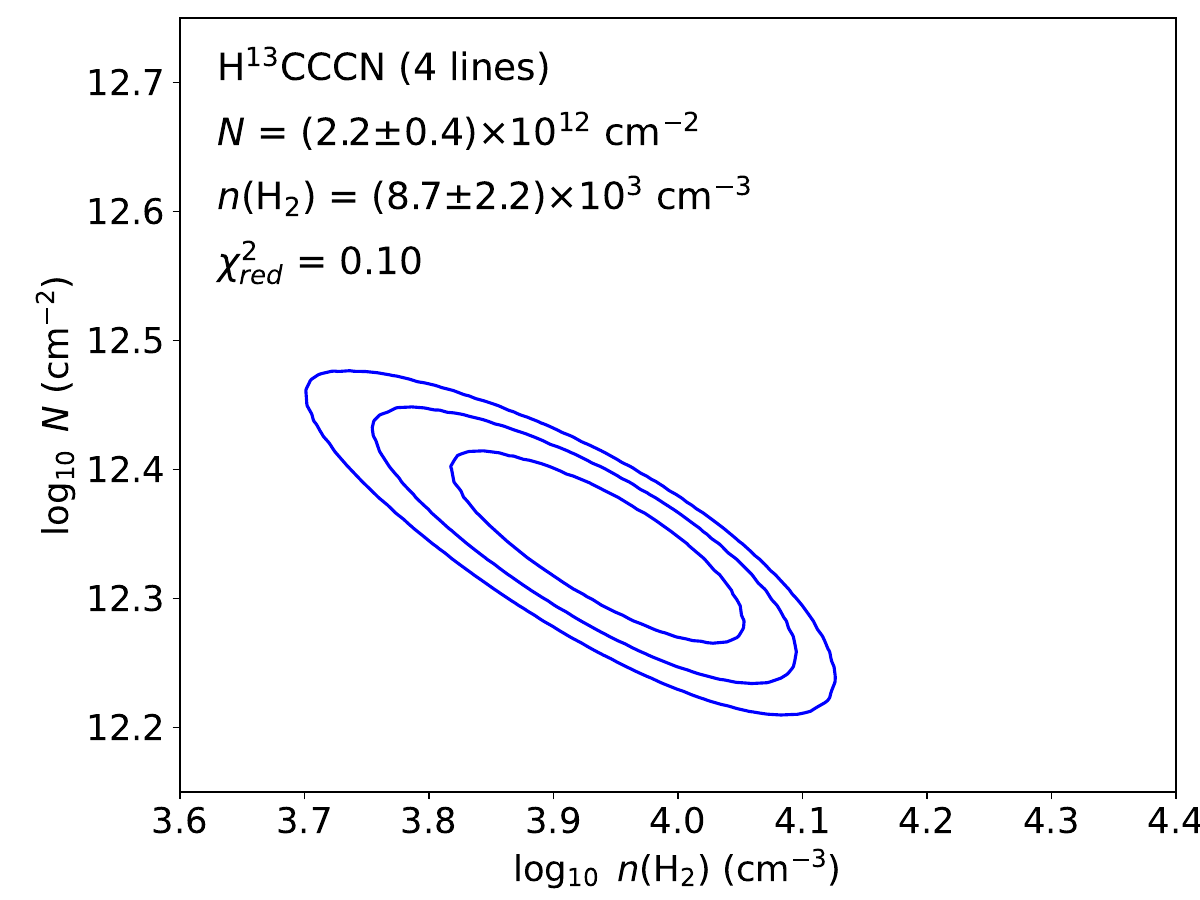} \hspace{0.05cm} \includegraphics[angle=0,width=0.325\textwidth]{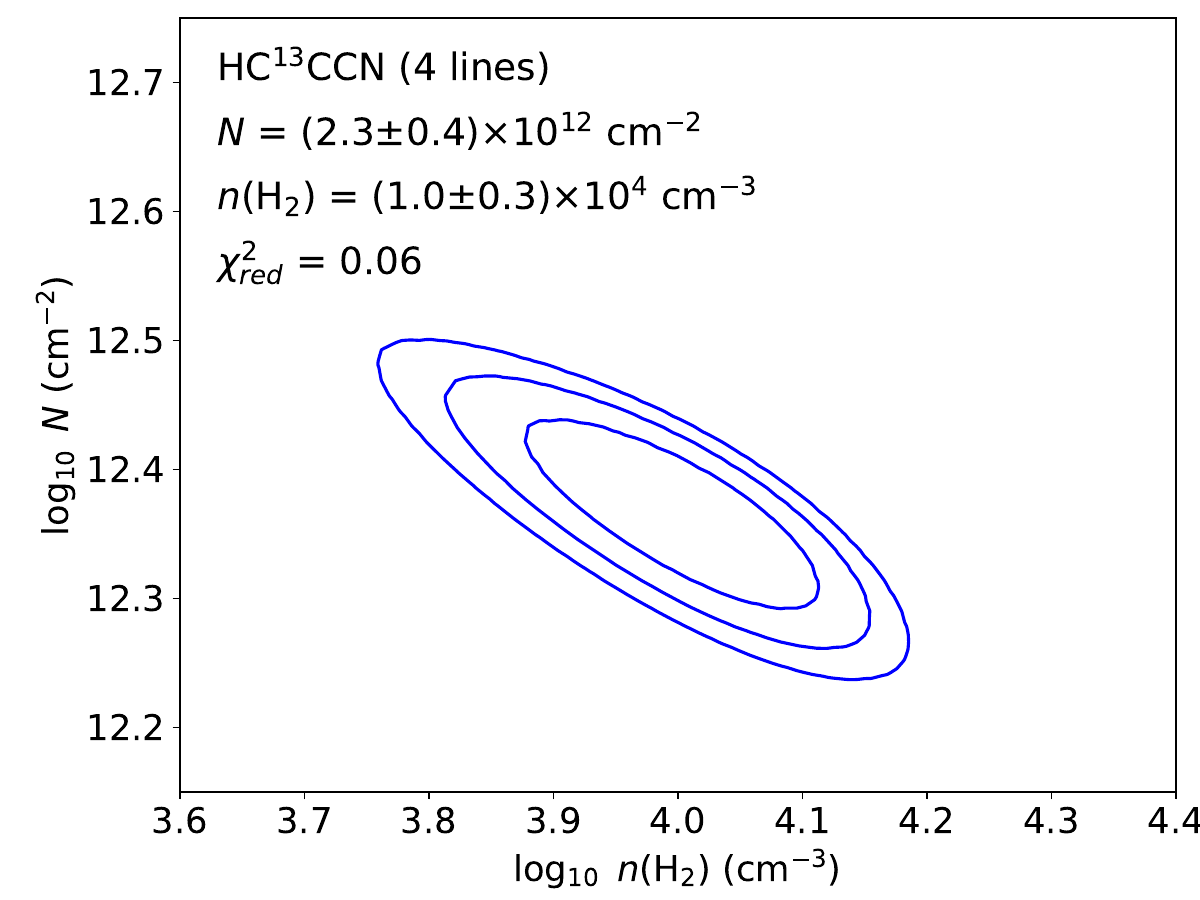} \hspace{0.05cm} \includegraphics[angle=0,width=0.325\textwidth]{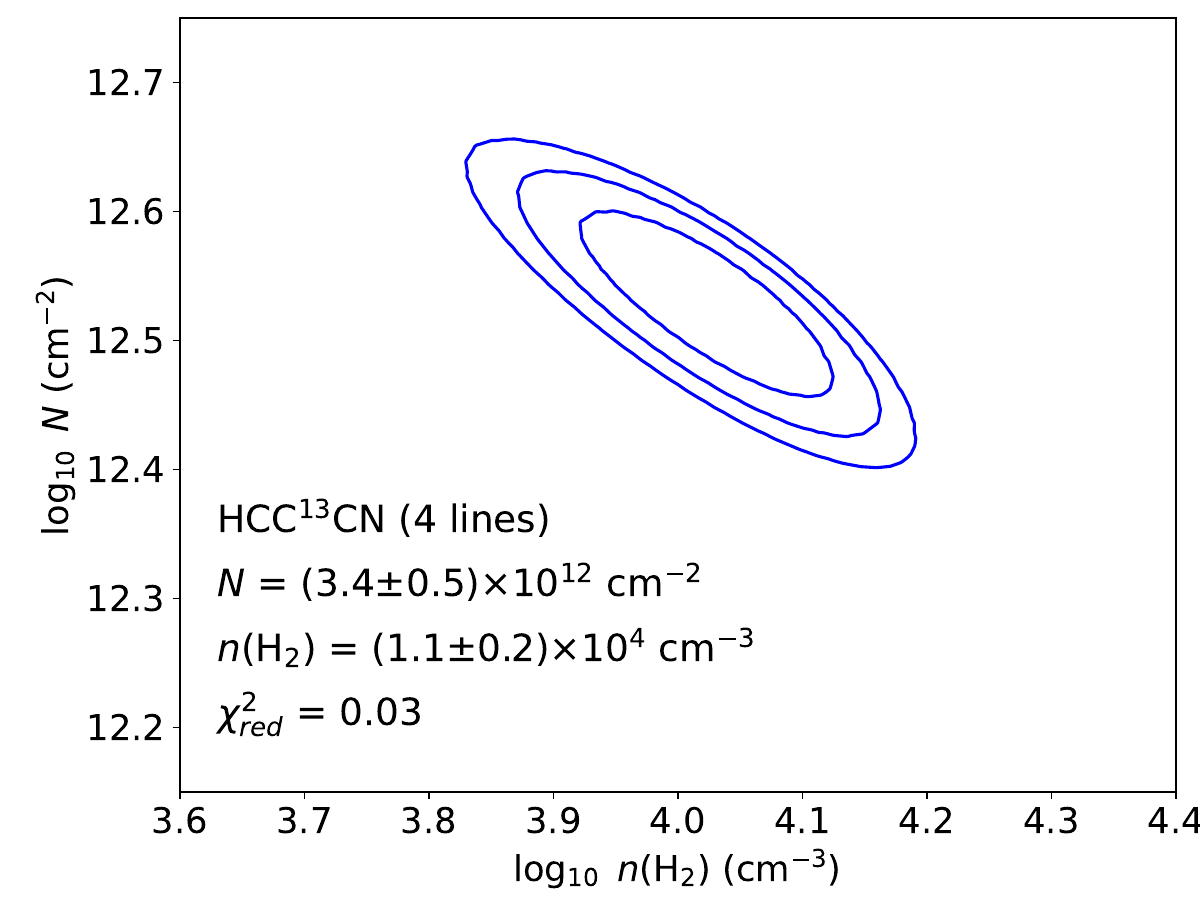}
\caption{$\chi^2$ as a function of H$_2$ volume density and column density of each of the three $^{13}$C isotopologues of HC$_3$N in TMC-1\,CP. Contours correspond to 1, 2, and 3\,$\sigma$ levels. The three maps have the same scale in the $x$ and $y$ axes to facilitate the comparison. The column density of HCC$^{13}$CN is clearly higher than those of H$^{13}$CCCN and HC$^{13}$CCN. The volume density of H$_2$ is constrained to a very narrow range, (0.9-1.1)\,$\times$\,10$^4$ cm$^{-3}$, by the three $^{13}$C isotopologues of HC$_3$N.}
\label{fig:hc3n_tmc1}
\end{figure*}

For TMC-1\,CP we derive kinetic temperatures of 8.8\,$\pm$\,0.6 K and 9.0\,$\pm$\,0.6 K from the $J$\,=\,5-4 and $J$\,=\,6-5 lines of CH$_3$CCH, respectively. Similarly, using the $J$\,=\,8-7 through $J$\,=\,12-11 lines of CH$_3$C$_4$H, which lie in the $Q$ band, we derive temperatures of 9.1\,$\pm$\,0.7 K, 8.7\,$\pm$\,0.6 K, 9.0\,$\pm$\,0.6 K, 8.1\,$\pm$\,0.7 K, and 9.1\,$\pm$\,0.8 K, respectively. We thus adopt a gas kinetic temperature of 9 K, which is slightly lower than values derived in previous studies, 11.0\,$\pm$\,1.0 K and 10.1\,$\pm$\,0.9 K at two positions close to the cyanopolyyne peak using NH$_3$ \citep{Feher2016} and 9.9\,$\pm$\,1.5 K from CH$_2$CCH \citep{Agundez2022}. In Lupus-1A we derive temperatures of 11.4\,$\pm$\,1.7 K and 10.2\,$\pm$\,1.1 K from the $J$\,=\,5-4 and $J$\,=\,6-5 lines of CH$_3$CCH, respectively. We thus adopt a gas kinetic temperature of 11 K, which is somewhat below the value of 14\,$\pm$\,2 K derived in \cite{Agundez2015} using the $K$\,=\,0, 1, and 2 components of the $J$\,=\,5-4 transition of CH$_3$CCH. In L1527 we derive 13.6\,$\pm$\,2.5 K and 15.1\,$\pm$\,2.4 K from the line parameters of CH$_3$CCH $J$\,=\,5-4 and $J$\,=\,6-5 reported by \cite{Yoshida2019}. We thus adopt a kinetic temperature of 14 K, which agrees perfectly with the value of 13.9 K derived by \cite{Sakai2008} using CH$_3$CCH as well. The gas kinetic temperature in L483 has been estimated to be 10 K by \cite{Anglada1997} using NH$_3$, while \cite{Agundez2019} derive values of 10 K and 15\,$\pm$\,2 K using either $^{13}$CO or CH$_3$CCH. A new analysis of the CH$_3$CCH data of \cite{Agundez2019} in which the weak $K$\,=\,3 components are neglected and only the $K$\,=\,1 and $K$\,=\,2 components are used results in kinetic temperatures of 11.5\,$\pm$\,1.1 K and 12.6\,$\pm$\,1.5 K, depending on whether the $J$\,=\,5-4 or $J$\,=\,6-5 transition is used. We thus adopt a kinetic temperature of 12 K for L483. For L1495B we derive 9.1\,$\pm$\,0.9 K and 9.2\,$\pm$\,0.7 K from CH$_3$CCH $J$\,=\,5-4 and $J$\,=\,6-5, and we thus adopt a kinetic temperature of 9 K. In L1521F we also adopt a gas kinetic temperature of 9 K since the derived temperatures from CH$_3$CCH $J$\,=\,5-4 and $J$\,=\,6-5 are 9.0\,$\pm$\,0.7 K and 8.9\,$\pm$\,0.9 K. The value agrees well with the temperature of 9.1\,$\pm$\,1.0 K derived by \cite{Codella1997} using NH$_3$. For the remaining cores, the gas kinetic temperatures were taken from the literature, as summarized in Table~\ref{table:sources}.

To estimate the volume density of H$_2$ we used the $^{13}$C isotopologues of HC$_3$N when these data were available. We have Yebes\,40m data of the $J$\,=\,4-3 and $J$\,=\,5-4 lines of H$^{13}$CCCN, HC$^{13}$CCN, and HCC$^{13}$CN for TMC-1\,CP, Lupus-1A, L1527, and L483. Data for one or various lines of these three isotopologues in the 3\,mm band are also available from the IRAM\,30m telescope (see Sect.~\ref{sec:observations}) or from the Nobeyama\,45 telescope (for L1527; see \citealt{Yoshida2019}). Using the $^{13}$C isotopologues of HC$_3$N turned out to constrain much better the H$_2$ density that using the main isotopologue because one gets rid of optical depth effects. We carried out non-LTE calculations under the Large Velocity Gradient (LVG) formalism adopting the gas kinetic temperature and linewidth given in Table~\ref{table:sources} and varying the column density of the $^{13}$C isotopologue of HC$_3$N and the H$_2$ volume density. As collision rate coefficients we used those calculated by \cite{Faure2016} for HC$_3$N with ortho and para H$_2$, where we adopted a low ortho-to-para ratio of H$_2$ of 10$^{-3}$, which is theoretically expected for cold dark clouds (e.g., \citealt{Flower2006}). The exact value of the ortho-to-para ratio of H$_2$ is not very important as long as the para form is well in excess of the ortho form, so that collisions with para H$_2$ dominate. The best estimates for the column density of the $^{13}$C isotopologue of HC$_3$N and the volume density of H$_2$ are found by minimizing $\chi^2$, which is defined as
\begin{equation}
\chi^2 = \sum_{i=1}^{N_l} \Bigg[ \frac{(I_{calc} - I_{obs})}{\sigma} \Bigg]^2,
\end{equation}
where the sum extends over the $N_l$ lines available, $I_{calc}$ and $I_{obs}$ are the calculated and observed velocity-integrated brightness temperatures, and $\sigma$ are the uncertainties in $I_{obs}$, which include the error given by the Gaussian fit and the calibration error of 10\,\%. To evaluate the goodness of the fit, we use the reduced $\chi^2$, which is defined as $\chi^2_{red}$\,=\,$\chi^2_{min}$/($N_l-p$), where $\chi^2_{min}$ is the minimum value of $\chi^2$ and $p$ is the number of free parameters. Typically, a value of $\chi^2_{red}$\,$\lesssim$\,1 indicates a good quality of the fit. In this case we have $p$\,=\,2 because there are two free parameters, the column density of the $^{13}$C isotopologue of HC$_3$N and the H$_2$ volume density. Errors in these two parameters are given as 1\,$\sigma$, where for $p$\,=\,2, the 1\,$\sigma$ level (68\,\% confidence) corresponds to $\chi^2$+2.3. The same statistical analysis is adopted in Sect.~\ref{sec:abundances} when studying molecular anions and their neutral counterparts through the LVG method. In some cases in which the number of lines is small or the H$_2$ density is poorly constrained, the H$_2$ volume density is kept fixed. In those cases $p$\,=\,1 and the 1\,$\sigma$ error (68\,\% confidence) in the column density is given by $\chi^2$+1.0.

In Fig.~\ref{fig:hc3n_tmc1} we show the results for TMC-1\,CP. In this starless core the H$_2$ volume density is well constrained by the four available lines of the three $^{13}$C isotopologues of HC$_3$N to a narrow range of (0.9-1.1)\,$\times$\,10$^4$ cm$^{-3}$ with very low values of $\chi^2_{red}$. We adopt as H$_2$ density in TMC-1\,CP the arithmetic mean of the values derived for the three isotopologues, i.e., 1.0\,$\times$\,10$^4$ cm$^{-3}$ (see Table~\ref{table:sources}). Similar calculations allow to derive H$_2$ volume densities of 1.8\,$\times$\,10$^4$ cm$^{-3}$ for Lupus-1A, 5.6\,$\times$\,10$^4$ cm$^{-3}$ for L483, and a lower limit of 10$^5$ cm$^{-3}$ for L1527 (see Table~\ref{table:sources}). The value for L483 is of the same order than those derived in the literature, 3.4\,$\times$\,10$^4$ cm$^{-3}$ from the model of \cite{Jorgensen2002} and 3\,$\times$\,10$^4$ cm$^{-3}$, from either NH$_3$ \citep{Anglada1997} or CH$_3$OH \citep{Agundez2019}. For L1495B we could only retrieve data for one of the $^{13}$C isotopologues of HC$_3$N, HCC$^{13}$CN, from which we derive a H$_2$ density of 1.6\,$\times$\,10$^4$ cm$^{-3}$ (see Table~\ref{table:sources}). In the case of L1521F, $^{13}$C isotopologues of HC$_3$N were not available and thus we used lines of HCCNC, adopting the collision rate coefficients calculated by \cite{Bop2021}, to derive a rough estimate of the H$_2$ volume density of 1\,$\times$\,10$^4$ cm$^{-3}$ (see Table~\ref{table:sources}). Higher H$_2$ densities, in the range (1-5)\,$\times$\,10$^5$ cm$^{-3}$, are derived for L1521F from N$_2$H$^+$ and N$_2$D$^+$ \citep{Crapsi2005}, probably because these molecules trace the innermost dense regions depleted in CO.

\begin{figure*}
\centering
\includegraphics[angle=0,width=\textwidth]{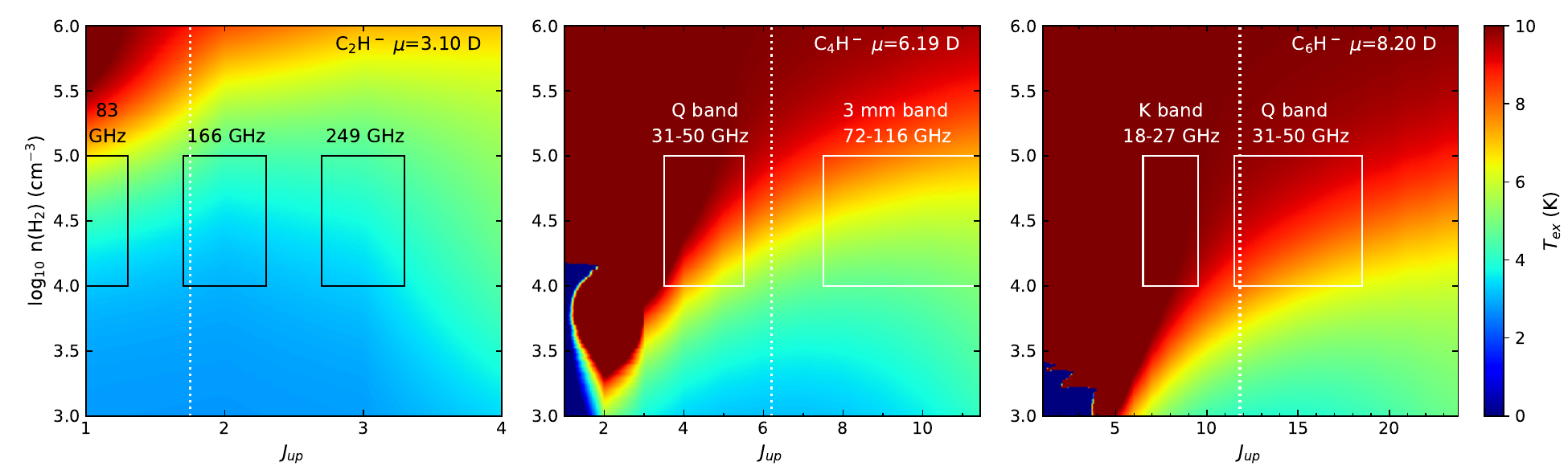} \includegraphics[angle=0,width=\textwidth]{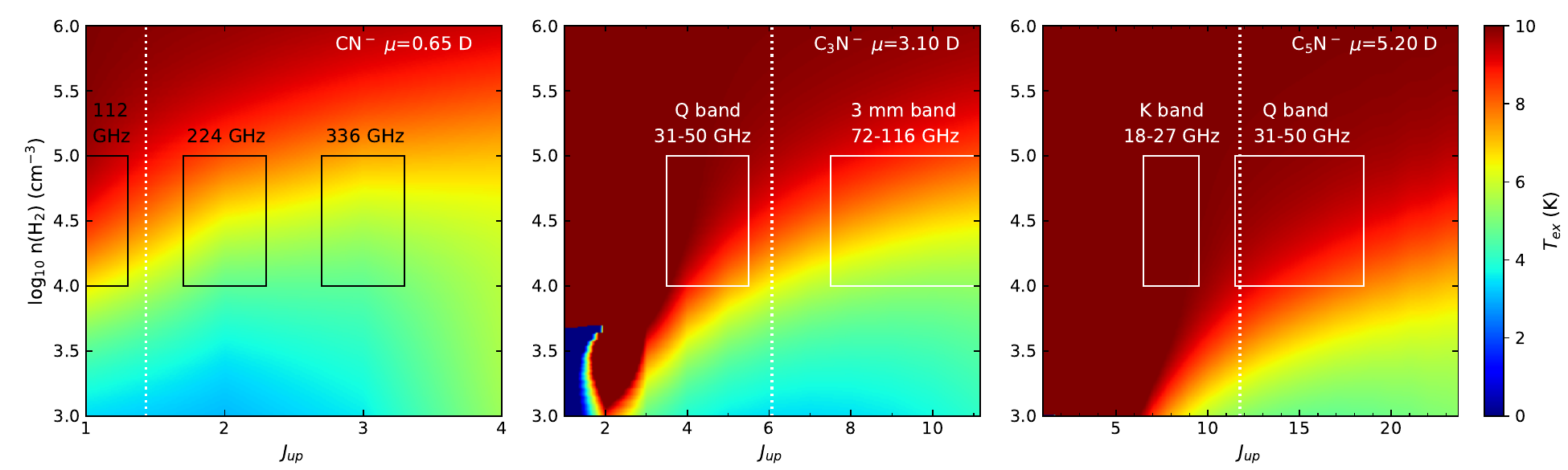}
\caption{Excitation temperature (color-coded map) as a function of quantum number of upper level (x-axis) and H$_2$ volume density (y-axis) for six negative molecular anions as obtained from LVG calculations adopting a gas kinetic temperature of 10 K, a column density of 10$^{11}$ cm$^{-2}$, and a linewidth of 0.5 km s$^{-1}$. The references for the dipole moments are \cite{Brunken2007b} for C$_2$H$^-$, \cite{Botschwina2000} for C$_4$H$^-$, \cite{Blanksby2001} for C$_6$H$^-$, \cite{Botschwina1995} for CN$^-$, \cite{Thaddeus2008} and \cite{Kolos2008} for C$_3$N$^-$, and \cite{Botschwina2008} for C$_5$N$^-$. For reference, the white dotted vertical line indicates the $J$ level at which the energy is 10 K. The microwave and mm spectral regions observable with radiotelescopes are indicated. The small dark blue regions in the bottom-left corner of the C$_4$H$^-$, C$_6$H$^-$, and C$_3$N$^-$ panels correspond to negative excitation temperatures.}
\label{fig:tex_anion}
\end{figure*}

For the remaining sources we adopted H$_2$ volume densities from the literature (see Table~\ref{table:sources}). For L1544 we adopted a value of 2\,$\times$\,10$^4$ cm$^{-3}$ from the analysis of SO and SO$_2$ lines by \cite{Vastel2018}. This H$_2$ density is in agreement with the range of values, (1.5-4.0)\,$\times$\,10$^4$ cm$^{-3}$, found by \cite{Bop2022} in their excitation analysis of HCCNC and HNC$_3$. Note that H$_2$ volume densities toward the dust peak are larger than 10$^6$ cm$^{-3}$. However, as shown by \cite{Spezzano2017}, the emission of C$_4$H probes the outer shells and thus a density of a few 10$^4$ cm$^{-3}$ is appropriate for our calculations toward the CH$_3$OH peak. In the cases of L1251A, L1512, L1172, L1389, and TMC-1\,C, we adopted the H$_2$ densities from the analysis of HC$_3$N lines by \cite{Cordiner2013}. The reliability of the H$_2$ volume densities derived by these authors is supported by the fact that the densities they derive for TMC-1\,CP and L1495B, 1.0\,$\times$\,10$^4$ cm$^{-3}$ and 1.1\,$\times$\,10$^4$ cm$^{-3}$, respectively, are close to the values determined in this study from $^{13}$C isotopologues of HC$_3$N (see Table~\ref{table:sources}).

In spite of the different evolutionary status of the 12 anion-containing clouds, the gas kinetic temperatures and H$_2$ volume densities at the scales proven by the Yebes\,40m, IRAM\,30m, and GBT telescopes are not that different. Gas temperatures are restricted to the very narrow range 9-14 K, while H$_2$ densities are in the range (1.0-7.5)\,$\times$\,10$^4$ cm$^{-3}$, at the exception of L1527 which has an estimated density in excess of 10$^5$ cm$^{-3}$ (see Table~\ref{table:sources}).

\section{Excitation of anions: general considerations} \label{sec:excitation}

One may expect that given the large dipole moments of molecular anions, as high as 10.4 D in the case of C$_8$H$^-$ \citep{Blanksby2001}, the rotational levels should be populated out of thermodynamic equilibrium in cold dark clouds. This is not always the case as it will be shown here. To get insight into the excitation of negative molecular ions in interstellar clouds we run non-LTE calculations under the LVG formalism adopting typical parameters of cold dark clouds, i.e., a gas kinetic temperature of 10 K, a column density of 10$^{11}$ cm$^{-2}$ (of the order of the values typically derived for anions in cold dark clouds; see references in Sect.~\ref{sec:observational_dataset}), and a linewidth of 0.5 km s$^{-1}$ (see Table~\ref{table:sources}), and we varied the volume density of H$_2$ between 10$^3$ and 10$^6$ cm$^{-3}$. The sets of rate coefficients for inelastic collisions with H$_2$ adopted are summarized in Table~\ref{table:collisions}. In those cases in which only collisions with He are available we scaled the rate coefficients by multiplying them by the square root of the ratio of the reduced masses of the H$_2$ and He colliding systems. When inelastic collisions for ortho and para H$_2$ are available, we adopted a ortho-to-para ratio of H$_2$ of 10$^{-3}$.

\begin{table}
\small
\caption{Collision rate coefficients used in this study.}
\label{table:collisions}
\centering
\begin{tabular}{llll}
\hline \hline
\multicolumn{1}{l}{Species} & \multicolumn{2}{l}{Collision data available$?$} & \multicolumn{1}{l}{Reference} \\
\multicolumn{1}{l}{} & \multicolumn{2}{l}{Adopted colliding system} & \multicolumn{1}{l}{} \\
\hline
\multicolumn{4}{c}{} \\
\multicolumn{4}{c}{Molecular anions} \\
\hline
C$_2$H$^-$ & Yes & C$_2$H$^-$ -- $p$-H$_2$ & \cite{Toumi2021} \\
C$_4$H$^-$ & Yes & C$_4$H$^-$ -- ($o$/$p$)-H$_2$ & \cite{Balanca2021} \\
C$_6$H$^-$ & Yes & C$_6$H$^-$ -- ($o$/$p$)-H$_2$ & \cite{Walker2017} \\
C$_8$H$^-$ & No  & C$_6$H$^-$ -- ($o$/$p$)-H$_2$ & \cite{Walker2017} \\
CN$^-$         & Yes & CN$^-$ -- ($o$/$p$)-H$_2$ & \cite{Klos2011} \\
C$_3$N$^-$ & Yes & C$_3$N$^-$ -- ($o$/$p$)-H$_2$ & \cite{Lara-Moreno2019} \\
C$_5$N$^-$ & No  & C$_6$H$^-$ -- ($o$/$p$)-H$_2$ & \cite{Walker2017} \\
\hline
\multicolumn{4}{c}{} \\
\multicolumn{4}{c}{Radicals} \\
\hline
C$_4$H & No  & HC$_3$N -- ($o$/$p$)-H$_2$ & \cite{Faure2016} \\
C$_6$H & Yes & C$_6$H -- He & \cite{Walker2018} \\
C$_8$H & No  & HC$_5$N -- $p$-H$_2$ & Lique (priv. comm.) \\
& & + IOS & \cite{Alexander1982} \\
C$_3$N & Yes & C$_3$N -- He & \cite{Lara-Moreno2021} \\
C$_5$N & No  & HC$_5$N -- $p$-H$_2$ & Lique (priv. comm.) \\
& & + IOS & \cite{Alexander1986} \\
\hline
\end{tabular}
\end{table}

In Fig.~\ref{fig:tex_anion} we show the calculated excitation temperatures ($T_{\rm ex}$) of lines of molecular anions as a function of the quantum number $J$ of the upper level and the H$_2$ volume density. The different panels correspond to different anions and show the regimes in which lines are either thermalized ($T_{\rm ex}$\,$\sim$\,10 K) of subthermally excited ($T_{\rm ex}$\,$<$\,10 K). To interpret these results it is useful to think in terms of the critical density, which for a given rotational level can be evaluated as the ratio of the de-excitation rates due to spontaneous emission and due to inelastic collisions (e.g., \citealt{Lara-Moreno2019}). Collision rate coefficients for transitions with $\Delta J$\,=\,$-$1 or $-$2, which are usually the most efficient, are of the order of 10$^{-10}$ cm$^3$ s$^{-1}$ at a temperature of 10 K for the anions for which calculations have been carried out (see Table~\ref{table:collisions}). The Einstein coefficient for spontaneous emission depends linearly on the square of the dipole moment and the cube of the frequency. Therefore, the critical density (and thus the degree of departure from LTE) is very different depending on the dipole moment of the anion and on the frequency of the transition. Regarding the dependence of the critical density on the dipole moment, C$_2$H$^-$ and CN$^-$ have a similar weight, and thus their low-$J$ lines, which are the ones observable for cold clouds, have similar frequencies. However, these two anions have quite different dipole moments, 3.1 and 0.65 Debye, respectively \citep{Brunken2007b,Botschwina1995}, which make them to show a different excitation pattern. As seen in Fig.~\ref{fig:tex_anion}, the low-$J$ lines of CN$^-$ are in LTE at densities above 10$^5$ cm$^{-3}$ while those of C$_2$H$^-$ require much higher H$_2$ densities to be in LTE. With respect to the dependence of the critical density with frequency, as one moves along the series of increasing weight C$_2$H$^-$ $\rightarrow$ C$_4$H$^-$ $\rightarrow$ C$_6$H$^-$ or CN$^-$ $\rightarrow$ C$_3$N$^-$ $\rightarrow$ C$_5$N$^-$ (see Fig.~\ref{fig:tex_anion}), the most favorable lines for detection in cold clouds (those with upper level energies around 10 K) shift to lower frequencies, which make the Einstein coefficients, and thus the critical densities, to decrease. That is, the lines of anions targeted by radiotelescopes are more likely to be thermalized for heavy anions than for light ones (see the higher degree of thermalization when moving from lighter to heavier anions in Fig.~\ref{fig:tex_anion}).

\begin{figure}
\centering
\includegraphics[angle=0,width=0.81\columnwidth]{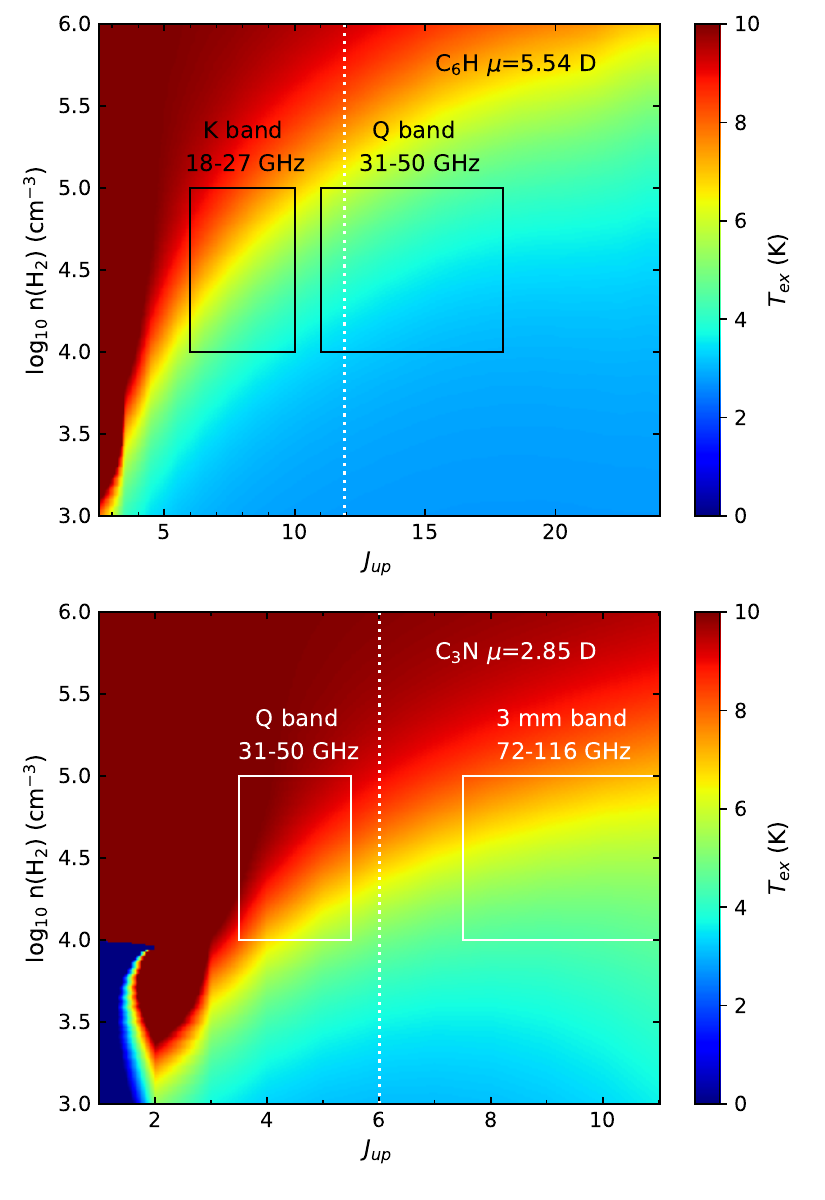}
\caption{Same as Fig.~\ref{fig:tex_anion} but for the radicals C$_6$H and C$_3$N adopting in this case a column density of 10$^{12}$ cm$^{-2}$. The references for the dipole moments are \cite{Woon1995} for C$_6$H and \cite{McCarthy1995} for C$_3$N.}
\label{fig:tex_neutral}
\end{figure}

The volume densities of H$_2$ in cold dark clouds are typically in the range 10$^4$-10$^5$ cm$^{-3}$ (see Table~\ref{table:sources}). Therefore, if C$_2$H$^-$ is detected in a cold dark cloud at some point in the future, the most favorable line for detection, the $J$\,=\,1-0, would be most likely subthermally excited, making necessary to use the collision rate coefficients to derive a precise abundance. In the case of a potential future detection of CN$^-$ in a cold interstellar cloud, the $J$\,=\,1-0 line would be in LTE only if the H$_2$ density of the cloud is $\geq$\,10$^5$ cm$^{-3}$ and out of LTE for lower densities (see Fig.~\ref{fig:tex_anion}). The medium-sized anions C$_4$H$^-$ and C$_3$N$^-$ are predicted to have their $Q$ band lines more or less close to LTE depending on whether the H$_2$ density is closer to 10$^5$ or to 10$^4$ cm$^{-3}$, while the lines in the 3\,mm band are likely to be subthermally excited unless the H$_2$ density is above 10$^5$ cm$^{-3}$ (see Fig.~\ref{fig:tex_anion}). For the heavier anions C$_6$H$^-$ and C$_5$N$^-$, the lines in the $K$ band are predicted to be thermalized at the gas kinetic temperature, while those in the $Q$ band may or may not be thermalized depending on the H$_2$ density (see Fig.~\ref{fig:tex_anion}). Comparatively, the $Q$ band lines of C$_5$N$^-$ are more easily thermalized than those of C$_6$H$^-$ because C$_5$N$^-$ has a smaller dipole moment than C$_6$H$^-$. We note that the results concerning C$_5$N$^-$ have to be taken with caution because we used the collision rate coefficients calculated for C$_6$H$^-$ in the absence of specific collision data for C$_5$N$^-$ (see Table~\ref{table:collisions}). We did similar calculations for C$_8$H$^-$, C$_{10}$H$^-$, and C$_7$N$^-$ (not shown) using the collision rate coefficients of C$_6$H$^-$. We find that the lines in a given spectral range deviate more from thermalization as the size of the anion increases. In the $K$ band, the lines of C$_6$H$^-$ and C$_5$N$^-$ are thermalized, while those of C$_{10}$H$^-$ become subthermally excited at low densities, around 10$^4$ cm$^{-3}$. In the $Q$ band the deviation from thermalization is even more marked for these large anions.

In summary, non-LTE calculations are particularly important to derive accurate abundances for anions when just one or two lines are detected and these lie in a regime of subthermal excitation, as indicated in Fig.~\ref{fig:tex_anion}. This becomes critical, in order of decreasing importance, for C$_2$H$^-$, CN$^-$, C$_4$H$^-$, C$_3$N$^-$, C$_6$H$^-$, C$_8$H$^-$, and C$_5$N$^-$ (for the three latter only if observed at frequencies above 30 GHz). The drawback is that the H$_2$ volume density must be known with a good precision if one aims at determining the anion column density accurately with only one or two lines.

\begin{table*}
\small
\caption{Results from LVG and rotation diagram analyses.}
\label{table:column_densities}
\centering
\begin{tabular}{llccccccccccl}
\hline \hline
\multicolumn{1}{l}{Source} & \multicolumn{1}{l}{Species} & \multicolumn{1}{l}{$N_l$\,$^a$} & & \multicolumn{1}{c}{$n$(H$_2$)} & \multicolumn{1}{c}{$N$} & \multicolumn{1}{c}{$\chi^2_{\rm red}$} & & \multicolumn{1}{c}{$T_{\rm rot}$} & \multicolumn{1}{c}{$N$} & & \multicolumn{1}{c}{$N$} & \multicolumn{1}{c}{Ref} \\
& & & & \multicolumn{1}{c}{(cm$^{-3}$)} & \multicolumn{1}{c}{(cm$^{-2}$)} & & & \multicolumn{1}{c}{(K)} & \multicolumn{1}{c}{(cm$^{-2}$)} & & \multicolumn{1}{c}{(cm$^{-2}$)} & \\
\cline{5-7} \cline{9-10} \cline{12-13}
& & & & \multicolumn{3}{c}{LVG} & & \multicolumn{2}{c}{Rotation diagram} & & \multicolumn{2}{c}{Literature} \\
\hline
TMC-1\,CP & C$_6$H$^-$ & 11 & & (5.9\,$\pm$\,1.6)\,$\times$\,10$^3$ & (1.5\,$\pm$\,0.2)\,$\times$\,10$^{11}$ & 0.38 & & 5.5\,$\pm$\,0.3 & (2.3\,$\pm$\,0.4)\,$\times$\,10$^{11}$ & & 1.0\,$\times$\,10$^{11}$ & (1) \\
TMC-1\,CP & C$_4$H$^-$ & 2 & & 5.9\,$\times$\,10$^3$\,$^b$ & (2.1\,$\pm$\,0.2)\,$\times$\,10$^{10}$ & -- & & 5.5\,$^b$ & 2.4\,$\times$\,10$^{10}$ & & 8.0\,$\times$\,10$^{9}$ & (2) \\
TMC-1\,CP & C$_8$H$^-$ & 12 & & (1.8\,$\times$\,0.8)\,$\times$\,10$^4$ & (2.0\,$\pm$\,0.4)\,$\times$\,10$^{10}$ & 0.97 & & 7.1\,$\pm$\,0.5 & (2.7\,$\pm$\,0.7)\,$\times$\,10$^{10}$ & & 2.1\,$\times$\,10$^{10}$ & (3) \\
TMC-1\,CP & C$_3$N$^-$ & 4 & & (1.5\,$\pm$\,0.6)\,$\times$\,10$^4$ & (6.4\,$\pm$\,0.8)\,$\times$\,10$^{10}$ & 0.04 & & 6.0\,$\pm$\,0.5 & (8.8\,$\pm$\,1.8)\,$\times$\,10$^{10}$ & & 1.3\,$\times$\,10$^{11}$ & (4) \\
TMC-1\,CP & C$_5$N$^-$ & 6 & & (5.4\,$\times$\,2.3)\,$\times$\,10$^3$ & (8.8\,$\times$\,1.4)\,$\times$\,10$^{10}$ & 0.44 & & 6.6\,$\pm$\,0.7 & (1.2\,$\pm$\,0.4)\,$\times$\,10$^{11}$ & & 2.6\,$\times$\,10$^{11}$ & (4) \\
Lupus-1A & C$_6$H$^-$ & 8 & & $>$\,1.5\,$\times$\,10$^4$ & (8.6\,$\pm$\,0.6)\,$\times$\,10$^{10}$ & 1.06 & & 12.0\,$\pm$\,1.5 & (8.8\,$\pm$\,1.5)\,$\times$\,10$^{10}$ & & 6.5\,$\times$\,10$^{10}$ & (5) \\
Lupus-1A & C$_4$H$^-$ & 4 & & (3.5\,$\pm$\,1.5)\,$\times$\,10$^4$ & (2.2\,$\pm$\,0.4)\,$\times$\,10$^{10}$ & 2.23 & & 6.9\,$\pm$\,0.9 & (3.2\,$\pm$\,0.9)\,$\times$\,10$^{10}$ & & 4.4\,$\times$\,10$^{10}$ & (5) \\
Lupus-1A & C$_8$H$^-$ & 2 & & 1.8\,$\times$\,10$^4$\,$^b$ & (1.9\,$\pm$\,0.3)\,$\times$\,10$^{10}$ & -- & & 12.0\,$^b$ & 2.1\,$\times$\,10$^{10}$ & & & \\
Lupus-1A & C$_3$N$^-$ & 1 & & 1.8\,$\times$\,10$^4$\,$^b$ & (4.0\,$\pm$\,1.2)\,$\times$\,10$^{10}$ & -- & & 12.0\,$^b$ & 5.1\,$\times$\,10$^{10}$ & & & \\
Lupus-1A & C$_5$N$^-$ & 5 & & $>$3\,$\times$\,10$^3$ & (5.5\,$\pm$\,0.8)\,$\times$\,10$^{10}$ & 1.52 & & 12.0\,$^b$ & 5.8\,$\times$\,10$^{10}$ & & & \\
L1527 & C$_6$H$^-$ & 8 & & $>$\,1\,$\times$\,10$^4$ & (4.5\,$\pm$\,0.5)\,$\times$\,10$^{10}$ & 1.23 & & 10.9\,$\pm$\,1.7 & (5.4\,$\pm$\,1.4)\,$\times$\,10$^{10}$ & & 5.8\,$\times$\,10$^{10}$ & (6) \\
L1527 & C$_4$H$^-$ & 4 & & $>$\,7\,$\times$\,10$^4$ & (1.5\,$\pm$\,0.2)\,$\times$\,10$^{10}$ & 0.18 & & 16.1\,$\pm$\,3.3 & (1.6\,$\pm$\,0.5)\,$\times$\,10$^{10}$ & & 1.6\,$\times$\,10$^{10}$ & (7) \\
L483 & C$_6$H$^-$ & 6 & & $>$1\,$\times$\,10$^4$ & (2.0\,$\pm$\,0.3)\,$\times$\,10$^{10}$ & 1.03 & & 12.0\,$^b$ & 2.1\,$\times$\,10$^{10}$ & & & \\
L483 & C$_4$H$^-$ & 2 & & 5.6\,$\times$\,10$^4$\,$^b$ & (6.4\,$\pm$\,1.3)\,$\times$\,10$^{9}$ & -- & & 12.0\,$^b$ & 8.9\,$\times$\,10$^{9}$ & & & \\
L1495B & C$_6$H$^-$ & 7 & & (2.3\,$\pm$\,1.0)\,$\times$\,10$^3$ & (4.5\,$\pm$\,1.7)\,$\times$\,10$^{10}$ & 1.62 & & 5.0\,$\pm$\,0.6 & (5.9\,$\pm$\,2.7)\,$\times$\,10$^{10}$ & & 3.4\,$\times$\,10$^{10}$ & (2) \\
L1544 & C$_6$H$^-$ & 5 & & $>$\,1\,$\times$\,10$^3$ & (2.5\,$\pm$\,1.0)\,$\times$\,10$^{10}$ & 1.01 & & 6.5\,$\pm$\,1.6 & (3.2\,$\pm$\,1.8)\,$\times$\,10$^{10}$ & & 3.1\,$\times$\,10$^{10}$ & (8) \\
L1521F & C$_6$H$^-$ & 1 & & 1\,$\times$\,10$^4$\,$^b$ & (3.4\,$\pm$\,0.8)\,$\times$\,10$^{10}$ & -- & & 9.0\,$^b$ & 4.2\,$\times$\,10$^{10}$ & & 3.4\,$\times$\,10$^{10}$ & (8) \\
L1251A & C$_6$H$^-$ & 2 & & 2.1\,$\times$\,10$^4$\,$^b$ & (2.2\,$\pm$\,0.4)\,$\times$\,10$^{10}$ & -- & & 10.0\,$^b$ & 2.5\,$\times$\,10$^{10}$ & & 2.3\,$\times$\,10$^{10}$ & (2) \\
L1512 & C$_6$H$^-$ & 2 & & 2.6\,$\times$\,10$^4$\,$^b$ & (1.4\,$\pm$\,0.2)\,$\times$\,10$^{10}$ & -- & & 10.0\,$^b$ & 1.6\,$\times$\,10$^{10}$ & & 1.5\,$\times$\,10$^{10}$ & (2) \\
L1172 & C$_6$H$^-$ & 2 & & 7.5\,$\times$\,10$^4$\,$^b$ & (2.4\,$\pm$\,0.3)\,$\times$\,10$^{10}$ & -- & & 10.0\,$^b$ & 2.5\,$\times$\,10$^{10}$ & & 2.4\,$\times$\,10$^{10}$ & (2) \\
L1389 & C$_6$H$^-$ & 2 & & 5.2\,$\times$\,10$^4$\,$^b$ & (2.0\,$\pm$\,0.3)\,$\times$\,10$^{10}$ & -- & & 10.0\,$^b$ & 2.2\,$\times$\,10$^{10}$ & & 2.1\,$\times$\,10$^{10}$ & (2) \\
TMC-1\,C & C$_6$H$^-$ & 2 & & 1.1\,$\times$\,10$^4$\,$^b$ & (4.5\,$\pm$\,0.5)\,$\times$\,10$^{10}$ & -- & & 10.0\,$^b$ & 5.2\,$\times$\,10$^{10}$ & & 4.8\,$\times$\,10$^{10}$ & (2) \\
& & & & & & & & \\
TMC-1\,CP & C$_6$H & 17 & & (7.5\,$\pm$\,2.8)\,$\times$\,10$^5$ & (4.8\,$\times$\,0.2)\,$\times$\,10$^{12}$ & 3.60 & & 7.0\,$\pm$\,0.3 & (6.2\,$\pm$\,0.7)\,$\times$\,10$^{12}$ & & 3.0\,$\times$\,10$^{12}$ & (6) \\
TMC-1\,CP & C$_4$H & 13 & & (8.6\,$\pm$\,0.9)\,$\times$\,10$^3$ & (8.5\,$\pm$\,0.7)\,$\times$\,10$^{13}$ & 3.29 & & 5.5\,$\pm$\,0.1 & (1.05\,$\pm$\,0.07)\,$\times$\,10$^{14}$ & & 7.1\,$\times$\,10$^{14}$ & (7) \\
TMC-1\,CP & C$_8$H & 21 & & 1.0\,$\times$\,10$^4$\,$^b$ & (3.0\,$\pm$\,0.1)\,$\times$\,10$^{11}$ & 2.86 & & 6.8\,$\pm$\,0.2 & (8.0\,$\pm$\,1.4)\,$\times$\,10$^{11}$ & & 4.6\,$\times$\,10$^{11}$ & (3) \\ 
TMC-1\,CP & C$_3$N & 10 & & (1.2\,$\pm$\,0.2)\,$\times$\,10$^4$ & (1.2\,$\pm$\,0.1)\,$\times$\,10$^{13}$ & 1.50 & & 4.8\,$\pm$\,0.1 & (1.7\,$\pm$\,0.2)\,$\times$\,10$^{13}$ & & 1.8\,$\times$\,10$^{13}$ & (4) \\
TMC-1\,CP & C$_5$N & 12 & & $>$\,1\,$\times$\,10$^3$ & (4.7\,$\pm$\,0.3)\,$\times$\,10$^{11}$ & 0.18 & & 9.1\,$\pm$\,0.9 & (4.8\,$\pm$\,1.0)\,$\times$\,10$^{11}$ & & 6.0\,$\times$\,10$^{11}$ & (4) \\ 
Lupus-1A & C$_6$H & 16 & & $>$\,7$\times$\,10$^5$ & (3.7\,$\pm$\,0.2)\,$\times$\,10$^{12}$ & 1.11 & & 10.7\,$\pm$\,0.7 & (3.8\,$\pm$\,0.4)\,$\times$\,10$^{12}$ & & 3.1\,$\times$\,10$^{12}$ & (5) \\
Lupus-1A & C$_4$H & 10 & & (1.2\,$\pm$\,0.2)\,$\times$\,10$^4$ & (8.4\,$\pm$\,0.9)\,$\times$\,10$^{13}$ & 1.56 & & 7.3\,$\pm$\,0.2 & (8.0\,$\pm$\,0.6)\,$\times$\,10$^{13}$ & & 5.0\,$\times$\,10$^{14}$ & (5) \\
Lupus-1A & C$_8$H & 2 & & 1.8\,$\times$\,10$^4$\,$^b$ & (2.7\,$\pm$\,0.4)\,$\times$\,10$^{11}$ & -- & & 10.7\,$^b$ & 2.8\,$\times$\,10$^{11}$ & & 3.5\,$\times$\,10$^{11}$ & (5) \\
Lupus-1A & C$_3$N & 8 & & (3.5\,$\pm$\,0.5)\,$\times$\,10$^4$ & (6.2\,$\pm$\,0.5)\,$\times$\,10$^{12}$ & 1.19 & & 6.8\,$\pm$\,0.2 & (8.1\,$\pm$\,0.8)\,$\times$\,10$^{12}$ & & & \\
Lupus-1A & C$_5$N & 10 & & 1.8\,$\times$\,10$^4$\,$^b$ & (3.1\,$\pm$\,0.2)\,$\times$\,10$^{11}$ & 1.38 & & 7.6\,$\pm$\,1.7 & (4.9\,$\pm$\,2.7)\,$\times$\,10$^{11}$ & & & \\ 
L1527      & C$_6$H & 16 & & $>$\,1.5\,$\times$\,10$^6$ & (8.8\,$\pm$\,0.4)\,$\times$\,10$^{11}$ & 0.66 & & 19.6\,$\pm$\,3.4 & (9.7\,$\pm$\,1.6)\,$\times$\,10$^{11}$ & & 6.2\,$\times$\,10$^{11}$ & (6) \\
L1527      & C$_4$H & 10 & & (1.4\,$\pm$\,0.6)\,$\times$\,10$^5$ & (2.9\,$\pm$\,0.1)\,$\times$\,10$^{13}$ & 0.29 & & 13.4\,$\pm$\,0.5 & (2.9\,$\pm$\,0.2)\,$\times$\,10$^{13}$ & & 1.5\,$\times$\,10$^{14}$ & (7) \\
L483        & C$_6$H & 14 & & (4.1\,$\pm$\,1.6)\,$\times$\,10$^5$ & (7.5\,$\pm$\,0.5)\,$\times$\,10$^{11}$ & 0.22 & & 8.3\,$\pm$\,0.6 & (8.7\,$\pm$\,1.5)\,$\times$\,10$^{11}$ & & \\
L483        & C$_4$H & 14 & & (1.3\,$\pm$\,0.2)\,$\times$\,10$^4$ & (2.3\,$\pm$\,0.2)\,$\times$\,10$^{13}$ & 3.94 & & 7.0\,$\pm$\,0.1 & (3.0\,$\pm$\,0.2)\,$\times$\,10$^{13}$ & & 1.2\,$\times$\,10$^{14}$ & (9) \\
L1495B   & C$_6$H & 16 & & (7.0\,$\pm$\,2.8)\,$\times$\,10$^5$ & (1.5\,$\pm$\,0.1)\,$\times$\,10$^{12}$ & 1.04 & & 7.0\,$\pm$\,0.3 & (1.8\,$\pm$\,0.2)\,$\times$\,10$^{12}$ & & 2.5\,$\times$\,10$^{12}$ & (2) \\
L1544     & C$_6$H & 11 & & (1.6\,$\pm$\,0.7)\,$\times$\,10$^5$ & (8.7\,$\pm$\,1.5)\,$\times$\,10$^{11}$ & 1.19 & & 5.4\,$\pm$\,0.4 & (1.4\,$\pm$\,0.3)\,$\times$\,10$^{12}$ & & 1.2\,$\times$\,10$^{12}$ & (8) \\
L1521F   & C$_6$H & 2 & & 1\,$\times$\,10$^4$\,$^b$ & (9.5\,$\pm$\,2.0)\,$\times$\,10$^{11}$ & -- & & 9.0\,$^b$ & 1.0\,$\times$\,10$^{12}$ & & 8\,$\times$\,10$^{11}$ & (8) \\
L1251A   & C$_6$H & 3 & & 2.1\,$\times$\,10$^4$\,$^b$ & (1.5\,$\pm$\,0.2)\,$\times$\,10$^{12}$ & -- & & 10.0\,$^b$ & 7.8\,$\times$\,10$^{11}$ & & 7.6\,$\times$\,10$^{11}$ & (2) \\
L1512     & C$_6$H & 5 & & 2.6\,$\times$\,10$^4$\,$^b$ & (7.6\,$\pm$\,0.6)\,$\times$\,10$^{11}$ & -- & & 10.0\,$^b$ & 5.5\,$\times$\,10$^{11}$ & & 4.6\,$\times$\,10$^{11}$ & (2) \\
L1172     & C$_6$H & 1 & & 7.5\,$\times$\,10$^4$\,$^b$ & (8.0\,$\pm$\,1.1)\,$\times$\,10$^{11}$ & -- & & 10.0\,$^b$ & 7.6\,$\times$\,10$^{11}$ & & 7.1\,$\times$\,10$^{11}$ & (2) \\
L1389     & C$_6$H & 3 & & 5.2\,$\times$\,10$^4$\,$^b$ & (4.4\,$\pm$\,0.6)\,$\times$\,10$^{11}$ & -- & & 10.0\,$^b$ & 5.0\,$\times$\,10$^{11}$  & & 4.7\,$\times$\,10$^{11}$ & (2) \\
TMC-1\,C & C$_6$H & 1 & & 1.1\,$\times$\,10$^4$\,$^b$ & (5.5\,$\pm$\,0.6)\,$\times$\,10$^{12}$ & -- & & 10.0\,$^b$ & 1.6\,$\times$\,10$^{12}$ & & 1.5\,$\times$\,10$^{12}$ & (2) \\
\hline
\end{tabular}
\tablenotec{\\
$^a$\,Number of lines included in the analysis.\\
$^b$\,Parameter was fixed to the value determined for a similar species, if possible, or to the value given in Table~\ref{table:sources}.\\
References: (1) \cite{McCarthy2006}. (2) \cite{Cordiner2013}. (3) \cite{Brunken2007a}. (4) \cite{Cernicharo2020}. (5) \cite{Sakai2010}. (6) \cite{Sakai2007}. (7) \cite{Agundez2008}. (8) \cite{Gupta2009}. (9) \cite{Agundez2019}.
}
\end{table*}

In the case of the neutral counterparts of molecular anions, collision rate coefficients have been calculated for C$_6$H and C$_3$N with He as collider \citep{Walker2018,Lara-Moreno2021}. We thus carried out LVG calculations similar to those presented before for anions. In this case we adopt a higher column density of 10$^{12}$ cm$^{-2}$, in line with typical values in cold dark clouds (see references in Sect.~\ref{sec:observational_dataset}). The results are shown in Fig.~\ref{fig:tex_neutral}. It is seen that in the case of C$_3$N, the excitation pattern is similar to that of the corresponding anion, C$_3$N$^-$, shown in Fig.~\ref{fig:tex_anion}. The thermalization of C$_3$N occurs at densities somewhat higher compared to C$_3$N$^-$, mainly because the collision rate coefficients calculated for C$_3$N with He \citep{Lara-Moreno2021} are smaller than those computed for C$_3$N$^-$ with para H$_2$ \citep{Lara-Moreno2019}.  We note that this conclusion may change if the collision rate coefficients of C$_3$N with H$_2$ are significantly larger than the factor of 1.39 due to the change in the reduced mass when changing He by H$_2$. In the case of C$_6$H however the excitation behavior is very different to that of C$_6$H$^-$ (compare C$_6$H$^-$ in Fig.~\ref{fig:tex_anion} with C$_6$H in Fig.~\ref{fig:tex_neutral}). The rotational levels of the radical are much more subthermally excited than those of the corresponding anion, with a difference in the critical density of about a factor of 30. This is a consequence of the much smaller collision rate coefficients calculated for C$_6$H with He \citep{Walker2018} compared to those calculated for C$_6$H$^-$ with para H$_2$ \citep{Walker2017}, a difference that is well beyond the factor of 1.40 due to the change in the reduced mass when changing He by H$_2$.

\section{Anion abundances} \label{sec:abundances}

\begin{figure*}
\centering
\includegraphics[angle=0,width=0.45\textwidth]{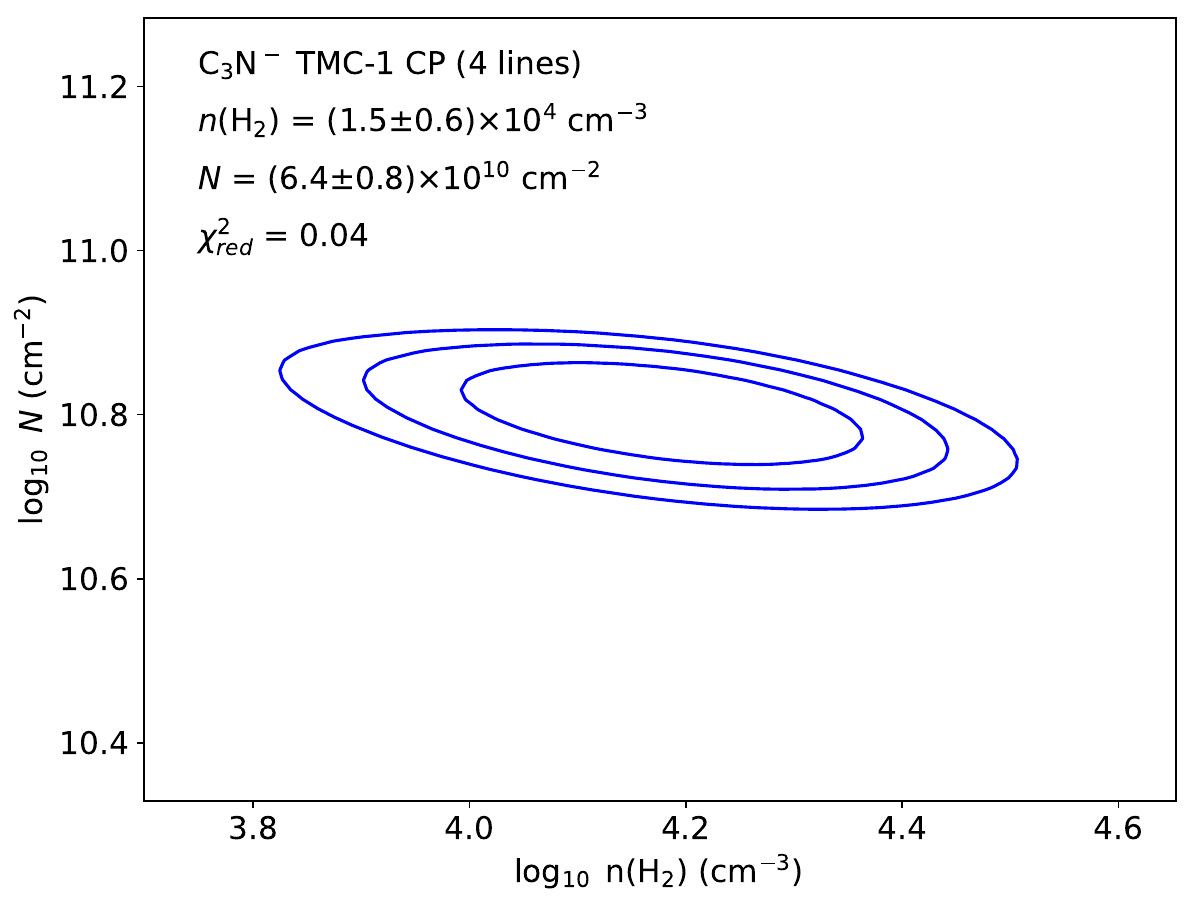} \hspace{0.50cm} \includegraphics[angle=0,width=0.45\textwidth]{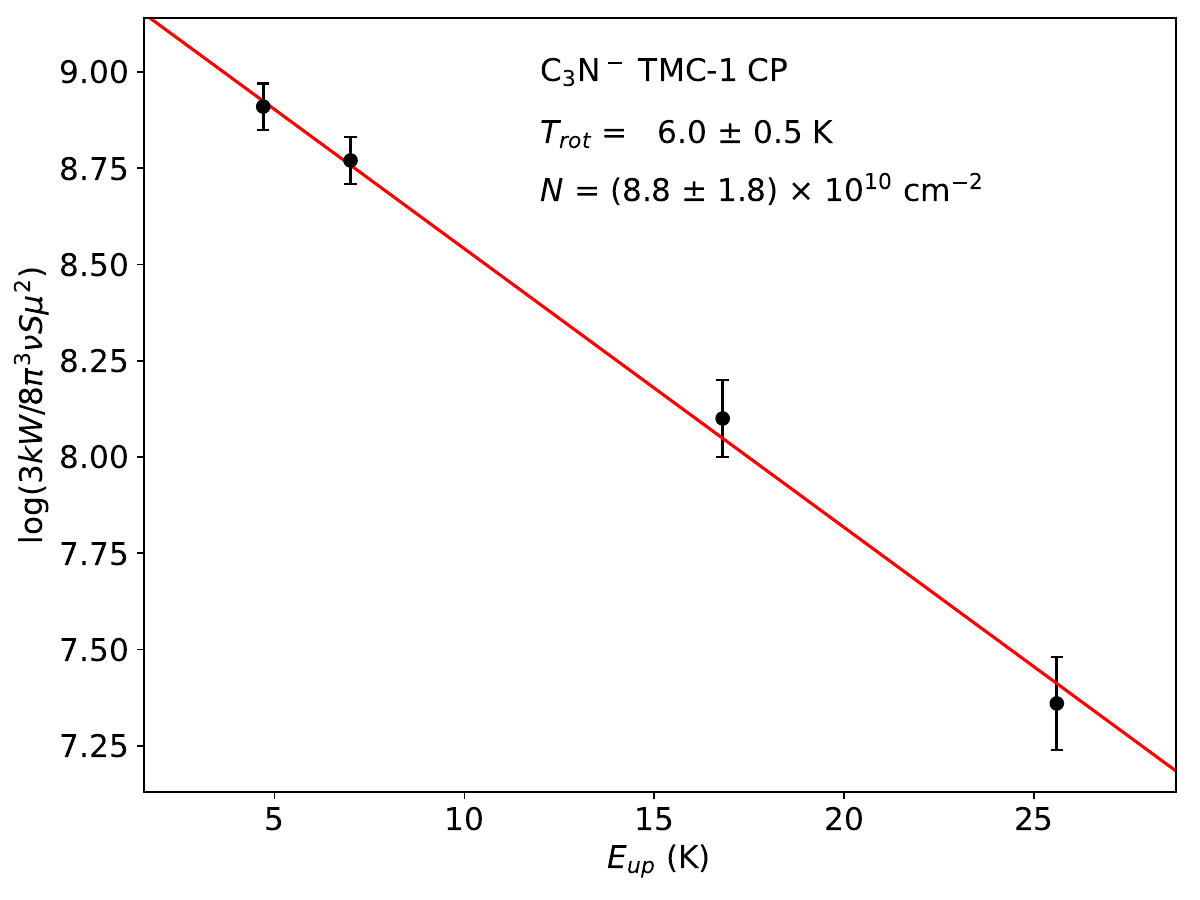}
\caption{Excitation and abundance analysis for C$_3$N$^-$ in TMC-1\,CP. The left panel shows $\chi^2$ as a function of the H$_2$ volume density and the column density of C$_3$N$^-$, where contours correspond to 1, 2, and 3\,$\sigma$ levels. The right panel shows the rotation diagram.}
\label{fig:c3n-_tmc1}
\end{figure*}

We evaluated the column densities of molecular anions and their corresponding neutral counterparts in the 12 studied sources by carrying out LVG calculations similar to those described in Sect.~\ref{sec:sources} for the $^{13}$C isotopologues of HC$_3$N. We used the collision rate coefficients given in Table~\ref{table:collisions}. Gas kinetic temperatures and linewidths were fixed to the values given in Table~\ref{table:sources}, the ortho-to-para ratio of H$_2$, when needed, was fixed to 10$^{-3}$, and both the column density of the species under study and the H$_2$ volume density were varied. The best estimates for these two parameters were found by minimization of $\chi^2$ (see Sect.~\ref{sec:sources}). In addition, to evaluate the rotational temperature, and thus the level of departure from LTE, and to have an independent estimate of the column density, we constructed rotation diagrams.

The LVG method should provide a more accurate determination of the column density than the rotation diagram, as long as the collision rate coefficients with para H$_2$ and the gas kinetic temperature are accurately known. If an independent determination of the H$_2$ volume density is available from some density tracer (in our case the $^{13}$C isotopologues of HC$_3$N are used in several sources), a good agreement between the values of $n$(H$_2$) obtained from the species under study and from the density tracer supports the reliability of the LVG analysis. We note that densities do not need to be similar if the species studied and the density tracer are distributed over different regions, although in our case we expect similar distributions for HC$_3$N, molecular anions, and their neutral counterparts, as long as all them are carbon chains. A low value of $\chi^2_{red}$, typically $\lesssim$\,1, is also indicative of the goodness of the LVG analysis. If the quality of the LVG analysis is not satisfactory or the collision rate coefficients are not accurate, a rotation diagram may still provide a good estimate of the column density if the number of detected lines is high enough and they span a wide range of upper level energies. Therefore, a high number of detected lines makes likely to end up with a correct determination of the column density. On the other hand, if only one or two lines are detected, the accuracy with which the column density can be determined relies heavily on whether the H$_2$ volume density, in the case of an LVG calculation, or the rotational temperature, in the case of the rotation diagram, are known with some confidence.

\begin{figure}[!hb]
\centering
\includegraphics[angle=0,width=\columnwidth]{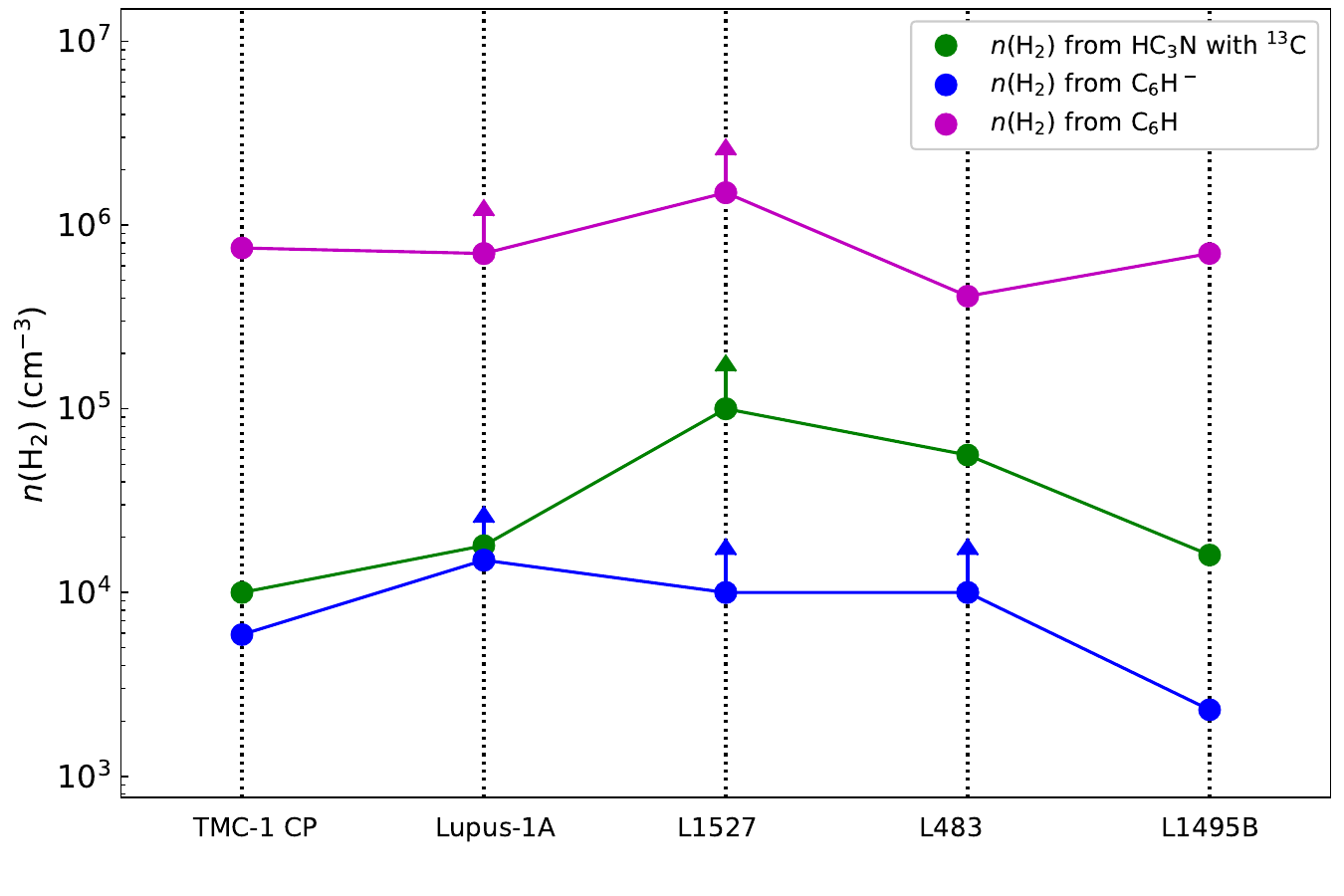}
\caption{Volume density of H$_2$ in various cold dark clouds determined through LVG calculations using different tracers. Green points correspond to densities derived from $^{13}$C isotopologues of HC$_3$N (see values in Table~\ref{table:sources}), while blue and magenta points correspond to densities obtained from the anion C$_6$H$^-$ and the radical C$_6$H, respectively (see values in Table~\ref{table:column_densities}). Note that H$_2$ densities derived through C$_6$H$^-$ are close to those derived by $^{13}$C isotopologues of HC$_3$N, while H$_2$ densities derived through C$_6$H are systematically higher by factors of 10-50.}
\label{fig:nh2}
\end{figure}

\begin{table*}
\small
\caption{Recommended column densities and anion-to-neutral abundance ratios.}
\label{table:recommended}
\centering
\begin{tabular}{lccccccc}
\hline \hline
 & \multicolumn{1}{c}{$N$ (cm$^{-2}$)} & \multicolumn{1}{c}{$N$ (cm$^{-2}$)} & & \multicolumn{1}{c}{Ratio ($\%$)} & & \multicolumn{1}{c}{Ratio ($\%$)} &  \multicolumn{1}{c}{Ref} \\
\hline
 & \multicolumn{4}{c}{This work} & & \multicolumn{2}{c}{Literature} \\
\cline{2-5} \cline{7-8}
& & & \\
 & \multicolumn{1}{c}{C$_4$H$^-$} & \multicolumn{1}{c}{C$_4$H} & & \multicolumn{3}{c}{C$_4$H$^-$/C$_4$H} & \\
\hline
TMC-1\,CP & (2.1\,$\pm$\,0.6)\,$\times$\,10$^{10}$ & (8.5\,$\pm$\,2.6)\,$\times$\,10$^{13}$ & & 0.025\,$\pm$\,0.007 & & 0.0012\,$\pm$\,0.0004 & (1) \\
Lupus-1A   & (2.2\,$\pm$\,0.7)\,$\times$\,10$^{10}$ & (8.4\,$\pm$\,1.3)\,$\times$\,10$^{13}$ & & 0.026\,$\pm$\,0.005 & & 0.0088\,$\pm$\,0.0053 & (2) \\
L1527        & (1.5\,$\pm$\,0.2)\,$\times$\,10$^{10}$ & (2.9\,$\pm$\,0.4)\,$\times$\,10$^{13}$ & & 0.052\,$\pm$\,0.004 & & 0.011 & (3) \\
L483          & (6.4\,$\pm$\,1.9)\,$\times$\,10$^{9}$ & (2.3\,$\pm$\,0.3)\,$\times$\,10$^{13}$ & & 0.028\,$\pm$\,0.008 & & & \\
& & & \\
 & \multicolumn{1}{c}{C$_6$H$^-$} & \multicolumn{1}{c}{C$_6$H} & & \multicolumn{3}{c}{C$_6$H$^-$/C$_6$H} & \\
\hline
TMC-1\,CP & (1.5\,$\pm$\,0.2)\,$\times$\,10$^{11}$ & (4.8\,$\pm$\,1.4)\,$\times$\,10$^{12}$ & & 3.1\,$\pm$\,0.6 & & 2.5 & (4) \\
Lupus-1A   & (8.6\,$\pm$\,1.3)\,$\times$\,10$^{10}$ & (3.7\,$\pm$\,1.1)\,$\times$\,10$^{12}$ & & 2.3\,$\pm$\,0.5 & & 2.1\,$\pm$\,0.6 & (2) \\
L1527        & (4.5\,$\pm$\,0.7)\,$\times$\,10$^{10}$ & (8.8\,$\pm$\,2.6)\,$\times$\,10$^{11}$ & & 5.1\,$\pm$\,1.1 & & 9.3\,$\pm$\,2.9 & (5) \\
L483          & (2.0\,$\pm$\,0.3)\,$\times$\,10$^{10}$ & (7.5\,$\pm$\,2.3)\,$\times$\,10$^{11}$ & & 2.7\,$\pm$\,0.6 & & & \\
L1495B     & (4.5\,$\pm$\,1.4)\,$\times$\,10$^{10}$ & (1.5\,$\pm$\,0.5)\,$\times$\,10$^{12}$ & & 3.0\,$\pm$\,0.8 & & 1.4\,$\pm$\,0.2 & (1) \\
L1544       & (2.5\,$\pm$\,0.8)\,$\times$\,10$^{10}$ & (8.7\,$\pm$\,2.6)\,$\times$\,10$^{11}$ & & 2.9\,$\pm$\,0.8 & & 2.5\,$\pm$\,0.8 & (6) \\
L1521F     & (3.4\,$\pm$\,1.0)\,$\times$\,10$^{10}$ & (1.0\,$\pm$\,0.3)\,$\times$\,10$^{12}$ & & 3.4\,$\pm$\,1.0 & & 4\,$\pm$\,1 & (6) \\
L1251A     & (2.2\,$\pm$\,0.7)\,$\times$\,10$^{10}$ & (7.8\,$\pm$\,2.3)\,$\times$\,10$^{11}$ & & 2.8\,$\pm$\,0.8 & & 3.0\,$\pm$\,0.6 & (1) \\
L1512       & (1.4\,$\pm$\,0.4)\,$\times$\,10$^{10}$ & (5.5\,$\pm$\,1.7)\,$\times$\,10$^{11}$ & & 2.5\,$\pm$\,0.7 & & 3.3\,$\pm$\,0.4 & (1) \\
L1172       & (2.4\,$\pm$\,0.7)\,$\times$\,10$^{10}$ & (7.6\,$\pm$\,2.3)\,$\times$\,10$^{11}$ & & 3.2\,$\pm$\,0.9 & & 3.3\,$\pm$\,0.5 & (1) \\
L1389       & (2.0\,$\pm$\,0.6)\,$\times$\,10$^{10}$ & (5.0\,$\pm$\,1.5)\,$\times$\,10$^{11}$ & & 4.0\,$\pm$\,1.1 & & 4.4\,$\pm$\,0.8 & (1) \\
TMC-1\,C & (4.5\,$\pm$\,1.4)\,$\times$\,10$^{10}$ & (1.6\,$\pm$\,0.5)\,$\times$\,10$^{12}$ & & 2.8\,$\pm$\,0.8 & & 3.1\,$\pm$\,0.3 & (1) \\
& & & \\
 & \multicolumn{1}{c}{C$_8$H$^-$} & \multicolumn{1}{c}{C$_8$H} & & \multicolumn{3}{c}{C$_8$H$^-$/C$_8$H} & \\
\hline
TMC-1\,CP & (2.0\,$\pm$\,0.3)\,$\times$\,10$^{10}$ & (3.0\,$\pm$\,0.9)\,$\times$\,10$^{11}$ & & 6.7\,$\pm$\,1.4 & & 5 & (7) \\
Lupus-1A   & (1.9\,$\pm$\,0.6)\,$\times$\,10$^{10}$ & (2.7\,$\pm$\,0.8)\,$\times$\,10$^{11}$ & & 7.0\,$\pm$\,2.0 & & 4.7\,$\pm$\,1.7 & (2) \\
& & & \\
 & \multicolumn{1}{c}{C$_3$N$^-$} & \multicolumn{1}{c}{C$_3$N} & & \multicolumn{3}{c}{C$_3$N$^-$/C$_3$N} & \\
\hline
TMC-1\,CP & (6.4\,$\pm$\,1.0)\,$\times$\,10$^{10}$ & (1.2\,$\pm$\,0.2)\,$\times$\,10$^{13}$ & & 0.53\,$\pm$\,0.04 & & 0.71 & (8) \\
Lupus-1A & (4.0\,$\pm$\,1.2)\,$\times$\,10$^{10}$ & (6.2\,$\pm$\,0.9)\,$\times$\,10$^{12}$ & & 0.65\,$\pm$\,0.13 & & & \\
& & & \\
 & \multicolumn{1}{c}{C$_5$N$^-$} & \multicolumn{1}{c}{C$_5$N} & & \multicolumn{3}{c}{C$_5$N$^-$/C$_5$N} & \\
\hline
TMC-1\,CP & (8.8\,$\pm$\,1.3)\,$\times$\,10$^{10}$ & (4.7\,$\pm$\,0.7)\,$\times$\,10$^{11}$ & & 19\,$\pm$\,1 & & 43 & (8) \\
Lupus-1A   & (5.5\,$\pm$\,0.8)\,$\times$\,10$^{10}$ & (3.1\,$\pm$\,0.5)\,$\times$\,10$^{11}$ & & 18\,$\pm$\,1 & & & \\
\hline
\end{tabular}
\tablenoted{\\
References: (1) \cite{Cordiner2013}. (2) \cite{Sakai2010}. (3) \cite{Agundez2008}. (4) \cite{McCarthy2006}. (5) \cite{Sakai2007}. (6) \cite{Gupta2009}. (7) \cite{Brunken2007a}. (8) \cite{Cernicharo2020}.
}
\end{table*}

In Table~\ref{table:column_densities} we present the results from the LVG analysis and the rotation diagram for all molecular anions detected in cold dark clouds and for the corresponding neutral counterparts, and compare the column densities derived with values from the literature, when available. In general, the column densities derived through the rotation diagram agree within 50\,\%, with those derived by the LVG analysis. The sole exceptions are C$_8$H in TMC-1\,CP and C$_6$H in TMC-1\,C. In the former case, the lack of specific collision rate coefficients for C$_8$H probably introduces an uncertainty in the determination of the column density. In the case of C$_6$H in TMC-1\,C, the suspected problem in the collision rate coefficients used for C$_6$H (see below) is probably behind the too large column density derived by the LVG method.

We first discuss the excitation and abundance analyses carried out for negative ions. For the anions detected in TMC-1\,CP through more than two lines, i.e., C$_6$H$^-$, C$_8$H$^-$, C$_3$N$^-$, and C$_5$N$^-$, the quality of the LVG analysis is good (in Fig.~\ref{fig:c3n-_tmc1} we show the case of C$_3$N$^-$). First, the number of lines available is sufficiently high and they cover a wide range of upper level energies. Second, the values of $\chi^2_{\rm red}$ are $\lesssim$\,1. And third, the H$_2$ densities derived are on the same order (within a factor of two) of that obtained through $^{13}$C isotopologues of HC$_3$N. The rotational temperatures derived by the rotation diagram indicate subthermal excitation, which is consistent with the H$_2$ densities derived and the excitation analysis presented in Sect.~\ref{sec:excitation}. We note that the column densities derived by the rotation diagram are systematically higher, by $\sim$\,50\,\%, compared to those derived through the LVG analysis. These differences are due to the breakdown of various assumptions made in the frame of the rotation diagram method, mainly the assumption of a uniform excitation temperature across all transitions and the validity of the Rayleigh-Jeans limit. Only the assumption that $\exp(h\nu/kT_{ex}) - 1$ = $h\nu/kT_{ex}$, implicitly made by the rotation diagram method in the Rayleigh-Jeans limit, already implies errors of 10-20\,\% in the determination of the column density for these anions. We therefore adopt as preferred values for the column densities those derived through the LVG method and assign an uncertainty of 15\,\%, which is the typical statistical error in the determination of the column density by the LVG analysis. The recommended values are given in Table~\ref{table:recommended}. Based on the same arguments, we conclude that the LVG analysis is satisfactory for C$_6$H$^-$ and C$_5$N$^-$ in Lupus-1A , C$_6$H$^-$ and C$_4$H$^-$ in L1527, and C$_6$H$^-$ in L483, and thus adopt the column densities derived by the LVG method with the same estimated uncertainty of 15\,\% (see Table~\ref{table:recommended}). In other cases the LVG analysis is less reliable due to a variety of reasons: only one or two lines are available (C$_4$H$^-$ in TMC-1\,CP, C$_8$H$^-$ and C$_3$N$^-$ in Lupus-1A, C$_4$H$^-$ in L483, and C$_6$H$^-$ in the clouds L1521F, L1251A, L1512, L1172, L1389, and TMC-1\,C), the parameter $\chi^2_{\rm red}$ is well above unity (C$_4$H$^-$ in Lupus-1A), or the column density has a sizable error (C$_6$H$^-$ in L1495B and L1544). In those cases we adopt the column densities derived by the LVG method but assign a higher uncertainty of 30\,\% (values are given in Table~\ref{table:recommended}).

In order to derive anion-to-neutral abundance ratios, we applied the same analysis carried out for the anions to the corresponding neutral counterparts. We first focus on the radical C$_6$H. There is one striking issue in the LVG analysis carried out for this species: the H$_2$ volume densities derived through C$_6$H are systematically higher, by 1-2 orders of magnitude, than those derived through the $^{13}$C isotopologues of HC$_3$N (see Fig.~\ref{fig:nh2}). This fact, together with the previous marked difference in the excitation pattern compared to that of C$_6$H$^-$ discussed in Sect.~\ref{sec:excitation}, suggests that the collision coefficients adopted for C$_6$H, which are based on the C$_6$H -- He system studied by \cite{Walker2018}, are too small. A further problem when using the collision coefficients of \cite{Walker2018} is that the line intensities from the $^2\Pi_{1/2}$ state, which in TMC-1\,CP are around 100 times smaller than those of the $^2\Pi_{3/2}$ state, are overestimated by a factor of $\sim$\,10. All these issues indicate that it is worth to undertake calculations of the collision rate coefficients of C$_6$H with H$_2$. The suspected problem in the collision rate coefficients of C$_6$H make us to adopt a conservative uncertainty of 30\,\% in the column densities derived. Moreover, in those sources in which C$_6$H is observed through just a few lines (L1521F, L1251A, L1512, L1172, L1389, and TMC-1\,C) we need to fix the H$_2$ density to the values derived through other density tracer (see Table~\ref{table:sources}), and given the marked difference between the H$_2$ densities derived through C$_6$H and other density tracers, it is likely that the C$_6$H column densities derived by the LVG method are unreliable. In these cases we therefore adopted as preferred C$_6$H column densities those obtained from the rotation diagram (see Table~\ref{table:recommended}). For the other neutral radicals, we adopted the column densities derived by the LVG method with an estimated uncertainty of 15\,\% when the LVG analysis was satisfactory (C$_3$N and C$_5$N in TMC-1\,CP, C$_4$H, C$_3$N, and C$_5$N in Lupus-1A, and C$_4$H in L1527) and a higher uncertainty of 30\,\% otherwise (C$_4$H and C$_8$H in TMC-1\,CP, C$_8$H in Lupus-1A, and C$_4$H in L483).

The recommended column densities for molecular anions and their neutral counterparts, and the corresponding anion-to-neutral ratios, are given in Table~\ref{table:recommended}. Since the lines of a given anion and its corresponding neutral counterpart where in most cases observed simultaneously, we expect the error due to calibration to cancel when computing anion-to-neutral ratios. We therefore subtracted the 10\,\% error due to calibration in the column densities when computing errors in the anion-to-neutral ratios. In general, the recommended anion-to-neutral abundance ratios agree within 50\,\% with the values reported in the literature, when available. Higher differences, of up to a factor of two, are found for C$_6$H$^-$ in L1527 and L1495B and for C$_5$N$^-$ in TMC-1\,CP. The most drastic differences are found for the C$_4$H$^-$/C$_4$H abundance ratio, for which we derive values much higher than those reported in the literature. The differences are largely due to the fact that here we adopt a revised value of the dipole moment of C$_4$H (2.10 D; \citealt{Oyama2020}), which is significantly higher than the value of 0.87 D calculated by \cite{Woon1995} and adopted in previous studies. This fact makes the column densities of C$_4$H to be revised downward by a factor of $\sim$\,6, and consequently the C$_4$H$^-$/C$_4$H ratios are also revised upward by the same factor.

\section{Discussion} \label{sec:discussion}

Having at hand a quite complete observational picture of negative ions in the interstellar medium, as summarized in Table~\ref{table:recommended}, it is interesting to examine which lessons can be learnt from this. There are at least two interesting aspects to discuss. First, how do the anion-to-neutral abundance ratio behave from one source to another, and whether the observed variations can be related to some property of the cloud. And second, within a given source, how do the anion-to-neutral abundance ratio vary for the different anions, and whether this can be related to the formation mechanism of anions.

Regarding the first point, since C$_6$H$^-$ is the most widely observed anion, it is very convenient to focus on it to investigate the source-to-source behavior of negative ions. The detection of C$_6$H$^-$ in L1527 and the higher C$_6$H$^-$/C$_6$H ratio derived in that source compared to that in TMC-1\,CP led \cite{Sakai2007} to suggest that this was a consequence of the higher H$_2$ density in L1527 compared to TMC-1\,CP. This point was later on revisited by \cite{Cordiner2013} with a larger number of sources detected in C$_6$H$^-$. These authors found a trend in which the C$_6$H$^-$/C$_6$H ratio increases with increasing H$_2$ density and further argued that this ratio increases as the cloud evolves from quiescent to star-forming, with ratios below 3\,\% in quiescent sources and above that value in star-forming ones.

There are theoretical grounds that support a relationship between the C$_6$H$^-$/C$_6$H ratio and the H$_2$ density. Assuming that the formation of anions is dominated by radiative electron attachment to the neutral counterpart and that they are mostly destroyed through reaction with H atoms, as expected for the conditions of cold dense clouds \citep{Flower2007}, it can be easily shown that at steady state the anion-to-neutral abundance ratio is proportional to the abundance ratio between electrons and H atoms, which in turn is proportional to the square root of the H$_2$ volume density (e.g., \citealt{Flower2007}). That is,
\begin{equation}
\rm \frac{C_6H^-}{C_6H} \propto \frac{e^-}{H} \propto {\textit n}(H_2)^{1/2}. \label{eq:c6hm_nh2}
\end{equation}

\begin{figure}
\centering
\includegraphics[angle=0,width=\columnwidth]{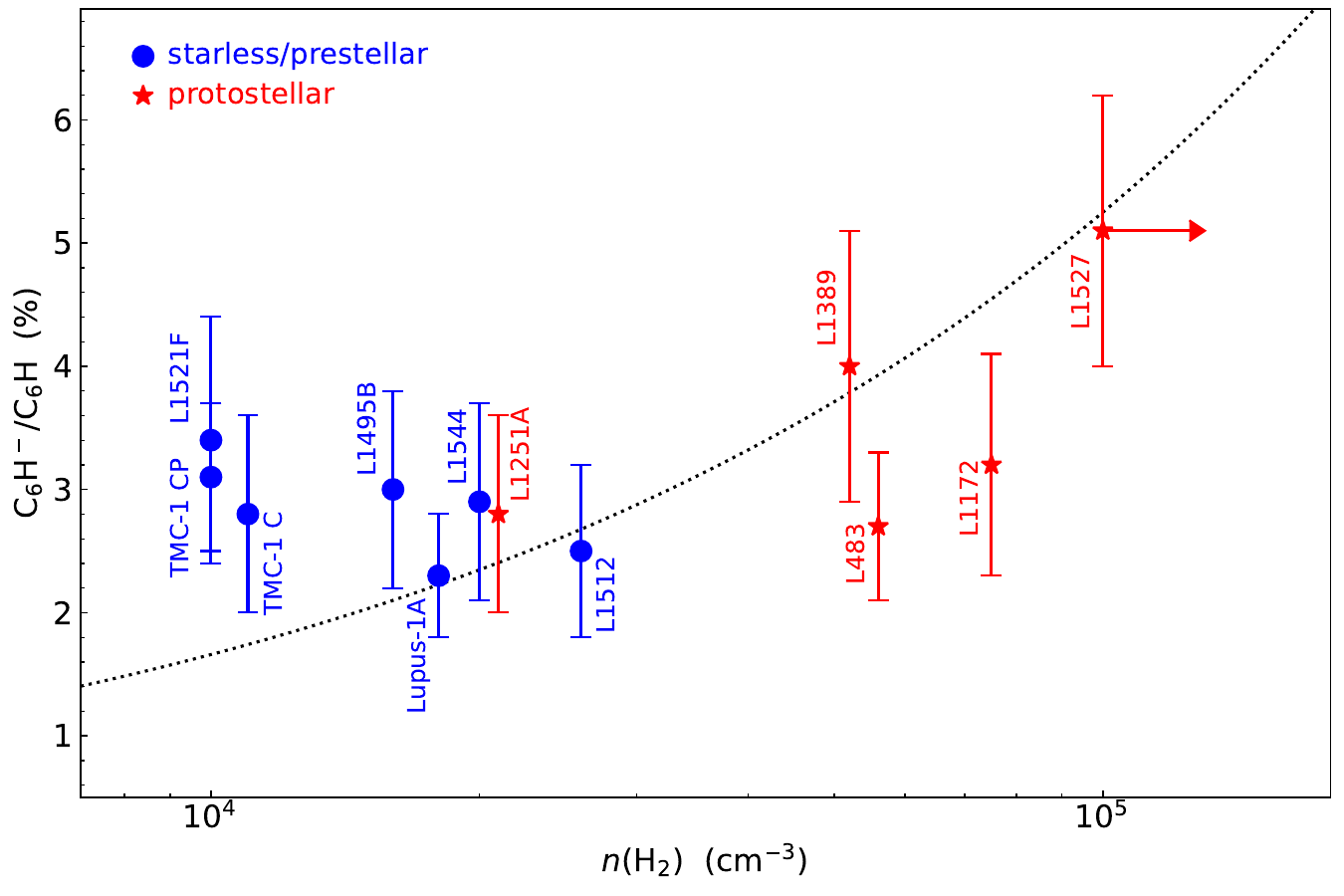}
\caption{Anion-to-neutral ratio C$_6$H$^-$/C$_6$H (see values in Table~\ref{table:recommended}) as a function of H$_2$ volume density. We do not give errors in the H$_2$ densities because this parameter is not derived in a coherent way for all sources (see Table~\ref{table:sources}). The dotted line represents the trend expected according to theory (see text).}
\label{fig:c6hm_nh2}
\end{figure}

\begin{figure*}
\centering
\includegraphics[angle=0,width=0.9\textwidth]{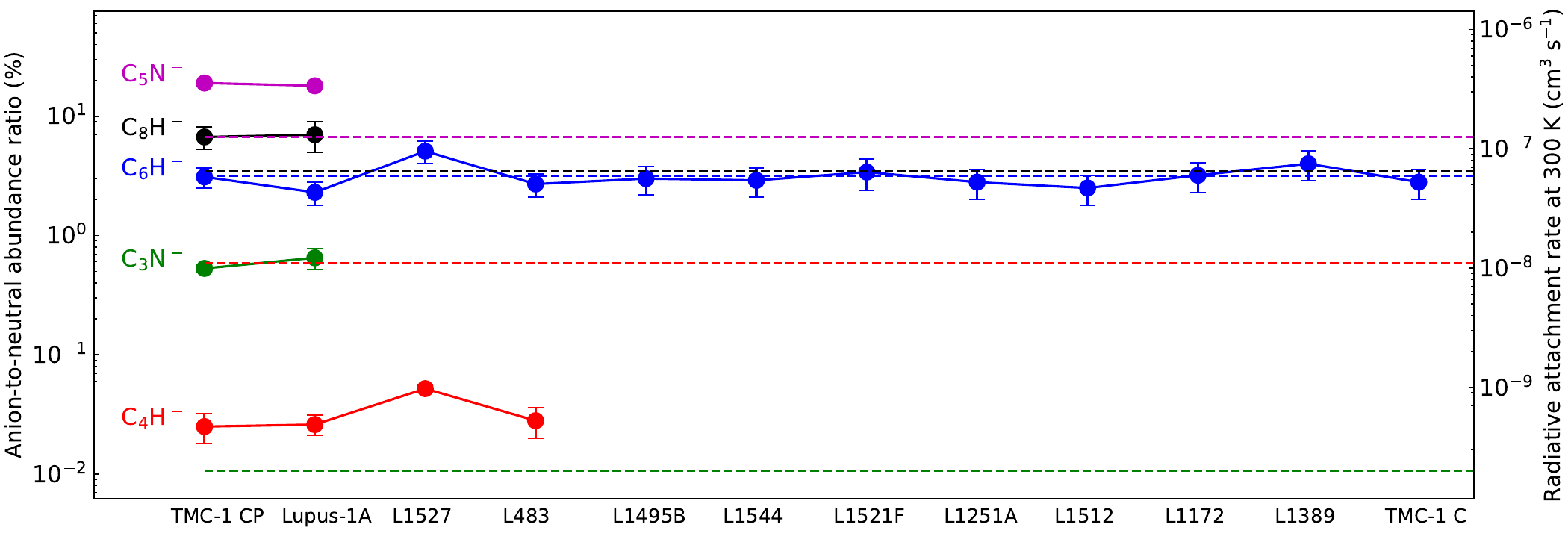}
\caption{Observed anion-to-neutral abundance ratios (referred to the left $y$ axis) in cold interstellar clouds where molecular anions have been detected to date. Values are given in Table~\ref{table:recommended}. Referred to the right $y$ axis and following the same color code we also plot as dotted horizontal lines the calculated rate coefficients at 300 K for the reaction of radiative electron attachment to the neutral counterpart. Adopted values are 1.1\,$\times$\,10$^{-8}$ cm$^3$ s$^{-1}$ for C$_4$H, 6.2\,$\times$\,10$^{-8}$ cm$^3$ s$^{-1}$ for C$_6$H and C$_8$H (\citealt{Herbst2008}; they are shown slightly displaced for visualization purposes), 2.0\,$\times$\,10$^{-10}$ cm$^3$ s$^{-1}$ for C$_3$N \citep{Petrie1997,Harada2008}, and 1.25\,$\times$\,10$^{-7}$ cm$^3$ s$^{-1}$ for C$_5$N \citep{Walsh2009}. The scale of the right $y$ axis is chosen to make the rate coefficient of electron attachment to C$_6$H to coincide with the mean of C$_6$H$^-$/C$_6$H ratios and to cover the same range in logarithmic scale than the left $y$ axis, which allows to visualize any potential proportionality between anion-to-neutral ratio and radiative electron attachment rate.}
\label{fig:anion2neutral}
\end{figure*}

In Fig.~\ref{fig:c6hm_nh2} we plot the observed C$_6$H$^-$/C$_6$H ratio as a function of the H$_2$ density for the 12 clouds where this anion has been detected. This is an extended and updated version of Figure~5 of \cite{Cordiner2013}, where we superimpose the theoretical trend expected according to Eq.~(\ref{eq:c6hm_nh2}). In general terms, the situation depicted by Fig.~\ref{fig:c6hm_nh2} is not that different from that found by \cite{Cordiner2013}. The main difference concerns L1495B, for which we derive a higher C$_6$H$^-$/C$_6$H ratio, 3.0\,\% instead of 1.4\,\%. Our value should be more accurate, given the larger number of lines used here. Apart from that, the C$_6$H$^-$/C$_6$H ratio tends to be higher in those sources with higher H$_2$ densities, which tend to be more evolved. This behavior is similar to that found by \cite{Cordiner2013}. The data points in Fig.~\ref{fig:c6hm_nh2} seem to be consistent with the theoretical expectation. We however caution that there is substantial dispersion in the data points. Moreover, the uncertainties in the anion-to-neutral ratios, together with those affecting the H$_2$ densities (not shown), make it difficult to end up with a solid conclusion on whether or not observations follow the theoretical expectations. If we restrict to the five best characterized sources (TMC-1\,CP, Lupus-1A, L1527, L483, and L1495B), all them observed in C$_6$H$^-$ through four or more lines and studied in the H$_2$ density in a coherent way, then the picture is such that all sources, regardless of its H$_2$ density, have similar C$_6$H$^-$/C$_6$H ratios, at the exception of L1527, which remains the only data point supporting the theoretical relation between anion-to-neutral ratio and H$_2$ density. It is also worth noting that when looking at C$_4$H$^-$, L1527 shows also an enhanced anion-to-neutral ratio compared to TMC-1\,CP, Lupus-1A, and L483. Further detections of C$_6$H$^-$ in sources with high H$_2$ densities, preferably above 10$^5$ cm$^{-3}$, should help to shed light on the suspected relation between anion-to-neutral ratio and H$_2$ density. This however may not be easy because chemical models predict that, although the C$_6$H$^-$/C$_6$H ratio increases with increasing H$_2$ density, an increase in the density also brings a decrease in the column density of both C$_6$H and C$_6$H$^-$ \citep{Cordiner2012}.

The second aspect that is worth to discuss is the variation of the anion-to-neutral ratio for different anions within a given source. Unlike the former source-to-source case, where variations were small (a factor of two at most), here anion-to-neutral ratios vary by orders of magnitude, i.e., well above uncertainties. Figure~\ref{fig:anion2neutral} summarizes the observational situation of interstellar anions in terms of abundances relative to their neutral counterpart. The variation of the anion-to-neutral ratios across different anions is best appreciated in TMC-1\,CP and Lupus-1A, which stand out as the two most prolific sources of interstellar anions. The lowest anion-to-neutral ratio is reached by far for C$_4$H$^-$, while the highest values are found for C$_5$N$^-$ and C$_8$H$^-$. We caution that the C$_5$N$^-$/C$_5$N ratio could have been overestimated if the true dipole moment of C$_5$N is a mixture between those of the $^2\Sigma$ and $^2\Pi$ states, as discussed by \cite{Cernicharo2008}, in a case similar to that studied for C$_4$H by \cite{Oyama2020}. For the large anion C$_7$N$^-$, the anion-to-neutral ratio is not known in TMC-1\,CP but it is probably large, as suggested by the detection of the lines of the anion and the non detection of the lines of the neutral \citep{Cernicharo2023a}. In the case of the even larger anion C$_{10}$H$^-$, the anion is found to be even more abundant than the neutral in TMC-1\,CP by a factor of two, although this result has probably an important uncertainty since the detection is done by line stack \citep{Remijan2023}. Moreover, it is yet to be confirmed that the species identified is C$_{10}$H$^-$ and not C$_9$N$^-$ \citep{Pardo2023}. In any case, a solid conclusion from the TMC-1\,CP and Lupus-1A data shown in Fig.~\ref{fig:anion2neutral} is that when looking at either the hydrocarbon series of anions or at the nitrile series, the anion-to-neutral ratio clearly increases with increasing size. The most straitforward interpretation of this behavior is related to the formation mechanism originally proposed by \cite{Herbst1981}, which relies on the radiative electron attachment (REA) to the neutral counterpart and for which the rate coefficient is predicted to increase markedly with increasing molecular size. 

If electron attachment is the dominant formation mechanism of anions and destruction rates are similar for all anions, we expect the anion-to-neutral abundance ratio to be proportional to the rate coefficient of radiative electron attachment. That is,
\begin{equation}
\rm \frac{A^-}{A} \propto {\textit k_{REA}}, \label{eq:rea}
\end{equation}
where A$^-$ and A are the anion and its corresponding neutral counterpart, respectively, and $k_{\rm REA}$ is the rate coefficient for radiative electron attachment to A.

To get insight into this relation we plot in Fig.~\ref{fig:anion2neutral} the rate coefficients calculated for the reactions of electron attachment forming the different anions on a scale designed on purpose to visualize if observed anion-to-neutral ratios scale with calculated electron attachment rates. We arbitrarily choose C$_6$H$^-$ as the reference for the discussion. If we first focus on the largest anion C$_8$H$^-$, we see that the C$_8$H$^-$/C$_8$H ratios are systematically higher, by a factor of 2-3, than the C$_6$H$^-$/C$_6$H ones, while \cite{Herbst2008} calculate identical electron attachment rates for C$_6$H and C$_8$H. Similarly, the C$_5$N$^-$/C$_5$N ratios are higher, by a factor of 6-8 than the C$_6$H$^-$/C$_6$H ratios, while the electron attachment rate calculated for C$_5$N is twice of that computed for C$_6$H in the theoretical scenario of \cite{Herbst2008}. That is, for the large anions C$_8$H$^-$ and C$_5$N$^-$ there is a deviation of a factor of 2-4 from the theoretical expectation given by Eq.~(\ref{eq:rea}). This deviation is small given the various sources of uncertainties in both the observed anion-to-neutral ratio (mainly due to uncertainties in the dipole moments) and the calculated electron attachment rate coefficient. The situation is different for the medium size anions C$_4$H$^-$ and C$_3$N$^-$. In the case of C$_4$H$^-$, anion-to-neutral ratios are $\sim$\,100 times lower than for C$_6$H$^-$, while the electron attachment rate calculated for C$_4$H is just $\sim$\,6 times lower than that computed for C$_6$H. The deviation from Eq.~(\ref{eq:rea}) of a factor $\sim$\,20, which is significant, is most likely due to the electron attachment rate calculated for C$_4$H by \cite{Herbst2008} being too large. In the case of C$_3$N$^-$, the observed anion-to-neutral ratios are 4-6 times lower than those derived for C$_6$H$^-$, while the electron attachment rate calculated by \cite{Petrie1997} for C$_3$N is 300 times lower than that computed for C$_6$H by \cite{Herbst2008}. Here the deviation is as large as two orders of magnitude and it is probably caused by the too low electron attachment rate calculated for C$_3$N.  In summary, calculated electron attachment rates are consistent with observed anion-to-neutral ratios for the large species but not for the medium-sized species C$_4$H and C$_3$N, in which cases calculated rates are too large by a factor of $\sim$\,20 and too small by a factor of $\sim$\,100, respectively.

Of course, the above conclusion holds in the scenario of anion formation dominated by electron attachment and similar destruction rates for all anions, which may not be strictly valid. For example, it has been argued \citep{Douguet2015,Khamesian2016,Forer2023} that the process of radiative electron attachment is much less efficient than calculated by \cite{Herbst2008}, with rate coefficients that are too small to sustain the formation of anions in interstellar space. \cite{Millar2017} discuss this point making the difference between direct and indirect radiative electron attachment, where for long carbon chains the direct process would be slow, corresponding to the rates calculated by \cite{Khamesian2016}, while the indirect process could be fast if a long-lived superexcited anion is formed, something that has some experimental support. \cite{Millar2017} conclude that there are enough grounds to support rapid electron attachment to large carbon chains, as calculated by \cite{Herbst2008}. The formation mechanism of anions through electron attachment is very selective for large species and thus has the advantage of naturally explaining the marked dependence of anion-to-neutral ratios with molecular size illustrated in Fig.~\ref{fig:anion2neutral}, something that would be difficult to explain through other formation mechanism. Indeed, mechanisms such as dissociative electron attachment to metastable isomers such as HNC$_3$ and H$_2$C$_6$ \citep{Petrie1997,Sakai2007} or reactions of H$^-$ with polyynes and cyanopolyynes \citep{Vuitton2009,Martinez2010,Khamesian2016,Gianturco2016,Murakami2022} could contribute to some extent but are unlikely to control the formation of anions since they can hardly explain why large anions are far more abundant than small ones.

\section{Conclusions}

We reported new detections of molecular anions in cold dense clouds and considerably expanded the number of lines through which negative ions are detected in interstellar clouds. The most prevalent anion remains to be C$_6$H$^-$, which to date has been seen in 12 interstellar clouds, while the rest of interstellar anions are observed in just 1-4 sources.

We carried out excitation calculations, which indicate that subthermal excitation is common for the lines of interstellar anions observed with radiotelescopes, with the low frequency lines of heavy anions being the easiest to thermalize. Important discrepancies between calculations and observations are found for the radical C$_6$H, which suggest that the collision rate coefficients currently available for this species need to be revisited.

We analyzed all the observational data acquired here and in previous studies through non-LTE LVG calculations and rotation diagrams to constrain the column density of each anion in each source. Differences in the anion-to-neutral abundance ratios with respect to literature values are small, less than 50\,\% in general and up to a factor of two in a few cases. The highest difference is found for the C$_4$H$^-$/C$_4$H ratio, which is shifted upward with respect to previous values due to the adoption of a higher dipole moment for the radical C$_4$H.

The observational picture of interstellar anions brought by this study shows two interesting results. On the one side, the C$_6$H$^-$/C$_6$H ratio seems to be higher in clouds with a higher H$_2$ density, which is usually associated to a later evolutionary status of the cloud, although error bars make it difficult to clearly distinguish this trend. On the other hand, there is a very marked dependence of the anion-to-neutral ratio with the size of the anion, which is in line with the formation scenario involving radiative electron attachment, the theory of which must still be revised for medium size species such as C$_4$H and C$_3$N.

\begin{acknowledgements}

We acknowledge funding support from Spanish Ministerio de Ciencia e Innovaci\'on through grants PID2019-106110GB-I00, PID2019-107115GB-C21, and PID2019-106235GB-I00.

\end{acknowledgements}

\appendix
\onecolumn

\section{Supplementary table}

\small
\begin{longtable}{lcc@{\hspace{0.05cm}}c@{\hspace{0.05cm}}cccc@{\hspace{0.05cm}}c@{\hspace{0.05cm}}ll}
\caption{\label{table:lines_anion} Observed line parameters of molecular anions in interstellar clouds.}\\
\hline \hline
\multicolumn{1}{l}{Species} & \multicolumn{1}{c}{Transition} & \multicolumn{1}{c}{Frequency} & & \multicolumn{1}{c}{$V_{\rm LSR}$} & \multicolumn{1}{c}{$\Delta v$} & \multicolumn{1}{c}{$T_A^*$ peak\,$^a$} & \multicolumn{1}{c}{$\int T_A^* dv$\,$^a$} & & Telescope & Reference \\
\multicolumn{1}{c}{} & \multicolumn{1}{c}{} & \multicolumn{1}{c}{(MHz)} & & \multicolumn{1}{c}{(km s$^{-1}$)} & \multicolumn{1}{c}{(km s$^{-1}$)} & \multicolumn{1}{c}{(mK)} & \multicolumn{1}{c}{(mK km s$^{-1}$)} & & & \\
\hline
\endfirsthead
\caption{continued.}\\
\hline\hline
\multicolumn{1}{l}{Species} & \multicolumn{1}{c}{Transition} & \multicolumn{1}{c}{Frequency} & & \multicolumn{1}{c}{$V_{\rm LSR}$} & \multicolumn{1}{c}{$\Delta v$} & \multicolumn{1}{c}{$T_A^*$ peak\,$^a$} & \multicolumn{1}{c}{$\int T_A^* dv$\,$^a$} & & Telescope & Reference \\
\multicolumn{1}{c}{} & \multicolumn{1}{c}{} & \multicolumn{1}{c}{(MHz)} & & \multicolumn{1}{c}{(km s$^{-1}$)} & \multicolumn{1}{c}{(km s$^{-1}$)} & \multicolumn{1}{c}{(mK)} & \multicolumn{1}{c}{(mK km s$^{-1}$)} & & & \\
\hline
\endhead
\hline
\endfoot
\hline \hline
\multicolumn{11}{c}{TMC-1\,CP} \\
\hline \hline
C$_6$H$^-$ &   4-3 & 11014.896 & & $+$5.80(2)  & 0.38(4) & 25(3) & 10.1(33) & & GBT & \cite{McCarthy2006} \\
           &   5-4 & 13768.614 & & $+$5.80(11) & 0.44(7) & 24(3) & 11.2(43) & & GBT & \cite{McCarthy2006} \\
           &  10-9 & 27537.130 & \multirow{2}{*}{\bigg\{} & & & & \multirow{2}{*}{41.6(90)\,$^{b, c}$} & \multirow{2}{*}{\bigg\}} & \multirow{2}{*}{GBT} & \multirow{2}{*}{\cite{Cordiner2013}} \\
           & 11-10 & 30290.813 & & & & & & & & \\
           & 12-11 & 33044.488 & & $+$5.78(1) & 0.73(1) & 22.3(23) & 17.4(18) & & Yebes\,40m & This work \\
           & 13-12 & 35798.153 & & $+$5.78(1) & 0.70(1) & 20.9(22) & 15.5(17) & & Yebes\,40m & This work \\
           & 14-13 & 38551.808 & & $+$5.78(1) & 0.64(2) & 18.9(20) & 12.8(14) & & Yebes\,40m & This work \\
           & 15-14 & 41305.453 & & $+$5.79(2) & 0.56(3) & 17.2(19) & 10.3(12) & & Yebes\,40m & This work \\
           & 16-15 & 44059.085 & & $+$5.79(2) & 0.57(3) & 12.8(15) &  7.7(10) & & Yebes\,40m & This work \\
           & 17-16 & 46812.706 & & $+$5.81(2) & 0.59(4) &  9.6(13) &  6.0(8)  & & Yebes\,40m & This work \\
           & 18-17 & 49566.313 & & $+$5.84(3) & 0.56(5) &  5.4(10) &  3.2(5)  & & Yebes\,40m & This work \\
\hline
C$_4$H$^-$ & 2-1 & 18619.761 & & $+$5.70(5)     & 0.43(13)   &        & 1.0(3)\,$^{b, d}$ & & GBT & \cite{Cordiner2013} \\ 
           & 4-3 & 37239.410 & & $+$5.81(2)     & 0.71(2)    & 6.0(7) & 4.5(6)            & & Yebes\,40m & This work \\
           & 5-4 & 46549.156 & & $+$5.81(2)     & 0.55(3)    & 5.8(8) & 3.4(4)            & & Yebes\,40m & This work \\
\hline
C$_8$H$^-$ & 11-10 & 12833.460 & & $+$5.71(5)  & 0.36(4)  &   8(1)   & 3.1(10)   & & GBT        & \cite{Brunken2007a} \\
           & 12-11 & 14000.134 & & $+$5.86(5)  & 0.37(4)  &   7(1)   & 2.8(10)   & & GBT        & \cite{Brunken2007a} \\
           & 13-12 & 15166.806 & & $+$5.84(6)  & 0.45(4)  &   6(1)   & 2.9(10)   & & GBT        & \cite{Brunken2007a} \\
           & 16-15 & 18666.814 & & $+$5.80(7)  & 0.34(5)  &  10(2)   & 3.6(16)  & & GBT        & \cite{Brunken2007a} \\
           & 27-26 & 31500.029 & & $+$5.82(4)  & 0.63(10) & 1.28(28) & 0.86(20) & & Yebes\,40m & This work \\
           & 28-27 & 32666.670 & & $+$5.76(3)  & 0.76(6)  & 1.08(26) & 0.87(15) & & Yebes\,40m & This work \\
           & 29-28 & 33833.309 & & $+$5.90(12) & 0.68(17) & 0.78(19) & 0.56(18) & & Yebes\,40m & This work \\
           & 30-29 & 34999.944 & & $+$5.86(6)  & 0.60(10) & 0.87(20) & 0.56(14) & & Yebes\,40m & This work \\
           & 31-30 & 36166.576 & & $+$5.83(8)  & 0.32(20) & 1.01(24) & 0.34(10) & & Yebes\,40m & This work \\
           & 32-31 & 37333.205 & & $+$5.73(5)  & 0.66(11) & 0.87(23) & 0.61(16) & & Yebes\,40m & This work \\
           & 33-32 & 38499.831 & & $+$5.81(9)  & 0.82(17) & 0.68(20) & 0.60(18) & & Yebes\,40m & This work \\
           & 34-33 & 39666.453 & & $+$5.93(10) & 0.40(12) & 0.44(21) & 0.19(7)\,$^e$  & & Yebes\,40m & This work \\ 
\hline
C$_3$N$^-$ &  4-3 & 38812.797 & & $+$5.78(1)  & 0.88(2)  & 4.2(2)  & 3.9(5)  & & Yebes\,40m & This work \\
           &  5-4 & 48515.872 & & $+$5.86(2)  & 0.61(4)  & 6.3(9)  & 4.1(6)  & & Yebes\,40m & This work \\
           &  8-7 & 77624.540 & & $+$5.88(3)  & 0.52(8)  & 7.1(17) & 3.9(9)  & & IRAM\,30m  & This work \\
           & 10-9 & 97029.687 & & $+$5.77(4)  & 0.38(6)  & 2.7(8)  & 1.1(3)  & & IRAM\,30m  & This work \\
\hline
C$_5$N$^-$ & 12-11 & 33332.570 & & $+$5.83(1) & 0.71(3) & 6.5(7) & 4.9(6) & & Yebes\,40m & This work \\
           & 13-12 & 36110.238 & & $+$5.80(1) & 0.64(2) & 6.1(7) & 4.1(5) & & Yebes\,40m & This work \\
           & 14-13 & 38887.896 & & $+$5.81(1) & 0.63(2) & 6.5(8) & 4.4(5) & & Yebes\,40m & This work \\
           & 15-14 & 41665.541 & & $+$5.82(2) & 0.58(2) & 5.7(7) & 3.5(5) & & Yebes\,40m & This work \\
           & 16-15 & 44443.173 & & $+$5.79(2) & 0.56(2) & 4.7(6) & 2.8(4) & & Yebes\,40m & This work \\
           & 17-16 & 47220.793 & & $+$5.81(2) & 0.50(4) & 3.6(6) & 1.9(3) & & Yebes\,40m & This work \\
\hline \hline
\multicolumn{11}{c}{Lupus-1A} \\
\hline \hline
C$_6$H$^-$ & 7-6   & 19276.037 & & $+$5.046(8)  & 0.16(2)  & 85(8)\,$^b$  & 14(2)\,$^b$ & & GBT        & \cite{Sakai2010} \\
           & 8-7   & 22029.741 & & $+$5.034(10) & 0.17(2)  & 94(11)\,$^b$ & 15(3)\,$^b$ & & GBT        & \cite{Sakai2010} \\
           & 12-11 & 33044.488 & & $+$5.06(2)   & 0.59(3)  & 30.1(37)     & 18.9(24)    & & Yebes\,40m & This work \\
           & 13-12 & 35798.153 & & $+$5.08(2)   & 0.51(3)  & 32.9(40)     & 17.8(25)    & & Yebes\,40m & This work \\
           & 14-13 & 38551.808 & & $+$5.05(2)   & 0.48(4)  & 30.4(38)     & 15.7(20)    & & Yebes\,40m & This work \\
           & 15-14 & 41305.453 & & $+$5.09(3)   & 0.40(7)  & 32.7(42)     & 13.8(19)    & & Yebes\,40m & This work \\
           & 16-15 & 44059.085 & & $+$5.07(3)   & 0.55(6)  & 24.2(35)     & 14.2(22)    & & Yebes\,40m & This work \\
           & 17-16 & 46812.706 & & $+$5.10(6)   & 0.51(8)  & 17.1(33)     & 9.3(18)     & & Yebes\,40m & This work \\
\hline
C$_4$H$^-$ & 4-3 & 37239.410 & & $+$5.078(13) & 0.34(3)  & 59(5)\,$^b$ & 19(5)\,$^b$ & & GBT        & \cite{Sakai2010} \\
           & 4-3 & 37239.410 & & $+$5.04(4)   & 0.78(7)  & 7.4(14)     & 6.1(11)     & & Yebes\,40m & This work \\
           & 5-4 & 46549.156 & & $+$5.05(9)   & 0.45(12) & 9.8(27)     & 4.7(13)     & & Yebes\,40m & This work \\
           & 9-8 & 83787.297 & & $+$5.23(6)   & 0.47(12) & 10.4(31)    & 5.3(13)     & & IRAM\,30m  & This work \\
\hline
C$_8$H$^-$ & 16-15   & 18666.814 & \multirow{2}{*}{\bigg\{} & \multirow{2}{*}{$+$5.014(11)} & \multirow{2}{*}{0.09(3)} & \multirow{2}{*}{35(9)} & \multirow{2}{*}{4(1)\,$^{b, c}$} & \multirow{2}{*}{\bigg\}} & \multirow{2}{*}{GBT} & \multirow{2}{*}{\cite{Sakai2010}} \\
           & 18-17 & 21000.145 & & & & & & & & \\
\hline
C$_3$N$^-$ &  4-3 & 38812.797 & & $+$5.16(15)  & 0.96(15)  & 2.8(10)  & 2.8(9)  & & Yebes\,40m & This work \\
\hline
C$_5$N$^-$ & 12-11 & 33332.570 & & $+$5.11(7)  & 0.50(9)  & 8.4(16) & 4.4(10) & & Yebes\,40m & This work \\
           & 13-12 & 36110.238 & & $+$5.11(7)  & 0.44(9)  & 6.5(13) & 3.1(7)  & & Yebes\,40m & This work \\
           & 14-13 & 38887.896 & & $+$5.13(7)  & 0.64(8)  & 8.0(17) & 5.4(11) & & Yebes\,40m & This work \\
           & 15-14 & 41665.541 & & $+$5.14(9)  & 0.37(10) & 9.2(19) & 3.7(9)  & & Yebes\,40m & This work \\
           & 16-15 & 44443.173 & & $+$5.09(10) & 0.58(15) & 6.1(18) & 3.8(11) & & Yebes\,40m & This work \\
\hline \hline
\multicolumn{11}{c}{L1527} \\
\hline \hline
C$_6$H$^-$ & 7-6   & 19276.037 & & $+$5.93(9) & 0.45(11) & 14(3)\,$^b$ & 7(2)\,$^b$  & & GBT        & \cite{Sakai2007} \\
           & 8-7   & 22029.741 & & $+$5.89(3) & 0.49(10) & 26(4)\,$^b$ & 18(4)\,$^b$ & & GBT        & \cite{Sakai2007} \\
           & 12-11 & 33044.488 & & $+$5.90(5) & 0.85(10) & 9.6(14)     & 8.6(16)     & & Yebes\,40m & This work \\
           & 13-12 & 35798.153 & & $+$5.85(4) & 0.60(4)  & 11.4(20)    & 7.3(18)     & & Yebes\,40m & This work \\
           & 14-13 & 38551.808 & & $+$5.84(3) & 0.61(5)  & 12.0(18)    & 7.8(12)     & & Yebes\,40m & This work \\
           & 15-14 & 41305.453 & & $+$5.90(3) & 0.60(4)  & 16.4(25)    & 10.4(19)    & & Yebes\,40m & This work \\
           & 16-15 & 44059.085 & & $+$5.90(3) & 0.52(4)  & 14.5(23)    & 8.0(16)     & & Yebes\,40m & This work \\
           & 17-16 & 46812.706 & & $+$5.83(5) & 0.58(8)  & 11.1(23)    & 6.8(14)     & & Yebes\,40m & This work \\
\hline
C$_4$H$^-$ & 4-3  & 37239.410 & & $+$5.92(12) & 0.80(20) & 3.2(10) & 2.7(7)  & & Yebes\,40m & This work \\
           & 5-4  & 46549.156 & & $+$6.05(15) & 0.73(15) & 4.9(19) & 3.8(13) & & Yebes\,40m & This work \\
           & 9-8  & 83787.297 & & $+$5.80(3)  & 0.62(9)  & 13(2)   & 8(1)    & & IRAM\,30m  & \cite{Agundez2008} \\
           & 10-9 & 93096.550 & & $+$5.90(4)  & 0.59(9)  & 11(2)   & 7(1)    & & IRAM\,30m  & \cite{Agundez2008} \\
\hline \hline
\multicolumn{11}{c}{L483} \\
\hline \hline
C$_6$H$^-$ & 12-11 & 33044.488 & & $+$5.38(6)  & 0.66(8)  & 4.9(11)  & 3.4(8)  & & Yebes\,40m & This work \\
           & 13-12 & 35798.153 & & $+$5.33(5)  & 0.70(7)  & 5.8(10)  & 4.3(8)  & & Yebes\,40m & This work \\
           & 14-13 & 38551.808 & & $+$5.33(5)  & 0.78(7)  & 5.2(9)   & 4.3(9)  & & Yebes\,40m & This work \\
           & 15-14 & 41305.453 & & $+$5.29(6)  & 0.46(9)  & 5.3(12)  & 2.6(6)  & & Yebes\,40m & This work \\
           & 16-15 & 44059.085 & & $+$5.24(10) & 0.75(12) & 4.8(12)  & 3.8(10) & & Yebes\,40m & This work \\
           & 17-16 & 46812.706 & & $+$5.34(7)  & 0.63(9)  & 5.0(14)  & 3.4(9)  & & Yebes\,40m & This work \\
\hline
C$_4$H$^-$ & 4-3 & 37239.410 & & $+$5.39(8)  & 0.73(12) & 2.8(7)  & 2.2(5) & & Yebes\,40m & This work \\
           & 5-4 & 46549.156 & & $+$5.37(10) & 0.44(15) & 2.7(12) & 1.3(5)\,$^e$ & & Yebes\,40m & This work \\ 
\hline \hline
\multicolumn{11}{c}{L1495B} \\
\hline \hline
C$_6$H$^-$ & 10-9  & 27537.130 & \multirow{2}{*}{\bigg\{} & & & & \multirow{2}{*}{9.6(20)\,$^{b, c}$} & \multirow{2}{*}{\bigg\}} & \multirow{2}{*}{GBT} & \multirow{2}{*}{\cite{Cordiner2013}} \\
           & 11-10 & 30290.813 & & & & & & & & \\
           & 12-11 & 33044.488 & & $+$7.66(5)  & 0.80(7)  & 5.9(12) & 5.0(9)  & & Yebes\,40m & This work \\
           & 13-12 & 35798.153 & & $+$7.65(5)  & 0.50(8)  & 5.8(12) & 3.1(6)  & & Yebes\,40m & This work \\
           & 14-13 & 38551.808 & & $+$7.58(7)  & 0.39(10) & 4.3(11) & 1.8(4)  & & Yebes\,40m & This work \\
           & 15-14 & 41305.453 & & $+$7.66(10) & 0.36(14) & 6.6(16) & 2.6(6)  & & Yebes\,40m & This work \\
           & 16-15 & 44059.085 & & $+$7.61(8)  & 0.49(12) & 4.1(11) & 2.1(6)  & & Yebes\,40m & This work \\
\hline \hline
\multicolumn{11}{c}{L1544} \\
\hline \hline
\multirow{2}{*}{C$_6$H$^-$} & \multirow{2}{*}{7-6} & \multirow{2}{*}{19276.037\,$^e$} & \multirow{2}{*}{\bigg\{} & $+$7.08(3) & 0.16(3) & 16(2) & \multirow{2}{*}{6.0(18)} & \multirow{2}{*}{\bigg\}} & \multirow{2}{*}{GBT} & \multirow{2}{*}{\cite{Gupta2009}} \\
                      &       &                   & & $+$7.30(3) & 0.13(3) & 26(2) &              & & & \\
           & 12-11 & 33044.488 & & $+$7.11(13) & 0.67(28) & 4.5(16) & 3.2(14) & & Yebes\,40m & This work \\
           & 13-12 & 35798.153 & & $+$7.04(10) & 0.48(16) & 4.1(12) & 2.1(9)  & & Yebes\,40m & This work \\
           & 14-13 & 38551.808 & & $+$6.98(8)  & 0.50(13) & 6.0(16) & 3.2(12) & & Yebes\,40m & This work \\
           & 15-14 & 41305.453 & & $+$7.34(18) & 0.76(36) & 4.6(15) & 3.7(16) & & Yebes\,40m & This work \\
\hline \hline
\multicolumn{11}{c}{L1521F} \\
\hline \hline
\multirow{2}{*}{C$_6$H$^-$} & \multirow{2}{*}{7-6} & \multirow{2}{*}{19276.037\,$^e$} & \multirow{2}{*}{\bigg\{} & $+$6.33(5) & 0.18(3) & 17(2) & \multirow{2}{*}{7.0(17)} & \multirow{2}{*}{\bigg\}} & \multirow{2}{*}{GBT} & \multirow{2}{*}{\cite{Gupta2009}} \\
                     &       &                   & & $+$6.64(5)   & 0.35(9) & 9(2) &              & & & \\
\hline \hline
\multicolumn{11}{c}{L1251A} \\
\hline \hline
C$_6$H$^-$ & 10-9   & 27537.130 & \multirow{2}{*}{\bigg\{} & & & & \multirow{2}{*}{6.5(17)\,$^{b, c}$} & \multirow{2}{*}{\bigg\}} & \multirow{2}{*}{GBT} & \multirow{2}{*}{\cite{Cordiner2013}} \\
                     & 11-10 & 30290.813 & & & & & & & & \\
\hline \hline
\multicolumn{11}{c}{L1512} \\
\hline \hline
C$_6$H$^-$ & 10-9   & 27537.130 & \multirow{2}{*}{\bigg\{} & & & & \multirow{2}{*}{4.3(8)\,$^{b, c}$} & \multirow{2}{*}{\bigg\}} & \multirow{2}{*}{GBT} & \multirow{2}{*}{\cite{Cordiner2013}} \\
                     & 11-10 & 30290.813 & & & & & & & & \\
\hline \hline
\multicolumn{11}{c}{L1172} \\
\hline \hline
C$_6$H$^-$ & 10-9   & 27537.130 & \multirow{2}{*}{\bigg\{} & & & & \multirow{2}{*}{6.7(15)\,$^{b, c}$} & \multirow{2}{*}{\bigg\}} & \multirow{2}{*}{GBT} & \multirow{2}{*}{\cite{Cordiner2013}} \\
           & 11-10 & 30290.813 & & & & & & & & \\
\hline \hline
\multicolumn{11}{c}{L1389} \\
\hline \hline
C$_6$H$^-$ & 10-9   & 27537.130 & \multirow{2}{*}{\bigg\{} & & & & \multirow{2}{*}{5.9(14)\,$^{b, c}$} & \multirow{2}{*}{\bigg\}} & \multirow{2}{*}{GBT} & \multirow{2}{*}{\cite{Cordiner2013}} \\
           & 11-10 & 30290.813 & & & & & & & & \\
\hline \hline
\multicolumn{11}{c}{TMC-1\,C} \\
\hline \hline
C$_6$H$^-$ & 10-9   & 27537.130 & \multirow{2}{*}{\bigg\{} & & & & \multirow{2}{*}{13.6(25)\,$^{b, c}$} & \multirow{2}{*}{\bigg\}} & \multirow{2}{*}{GBT} & \multirow{2}{*}{\cite{Cordiner2013}} \\
           & 11-10 & 30290.813 & & & & & & & & \\
\end{longtable}
\tablenotea{\\
$^a$\,Unless otherwise stated, the intensity scale is antenna temperature ($T_A^*$). It can be converted to main beam brightness temperature ($T_{\rm mb}$) by dividing by $B_{\rm eff}$/$F_{\rm eff}$, where $B_{\rm eff}$ is the main beam efficiency and $F_{\rm eff}$ is the telescope forward efficiency. For the Yebes\,40m telescope in the $Q$ band $B_{\rm eff}$\,=\,0.797\,$\exp$[$-$($\nu$(GHz)/71.1)$^2$] and $F_{\rm eff}$\,=\,0.97 (\texttt{https://rt40m.oan.es/rt40m\_en.php}), for the IRAM\,30m telescope $B_{\rm eff}$\,=\,0.871\,$\exp$[$-$($\nu$(GHz)/359)$^2$] and $F_{\rm eff}$\,=\,0.95 (\texttt{https://publicwiki.iram.es/Iram30mEfficiencies}), and for the GBT telescope we adopt $F_{\rm eff}$\,=\,1.0 and $B_{\rm eff}$\,=\,1.32\,$\times$\,0.71\,$\exp$[$-$($\nu$(GHz)/103.7)$^2$] \citep{Frayer2018}. The error in $\int T_A^* dv$ includes the contributions from the Gaussian fit and from calibration (assumed to be 10\,\%). $^b$\,Intensity scale is $T_{\rm mb}$. $^c$\,Average of two lines. $^d$\,Line neglected in the analysis. Intensity should be $\sim$\,3 times larger to be consistent with the other lines.
$^e$\,Line detected marginally.
}

\small
\begin{longtable}{llccll}
\caption{\label{table:lines_neutral} Observed velocity-integrated line intensities of neutral counterparts of molecular anions in interstellar clouds.}\\
\hline \hline
\multicolumn{1}{l}{Species} & \multicolumn{1}{c}{Transition} & \multicolumn{1}{c}{Frequency (MHz)} & \multicolumn{1}{c}{$\int T_A^* dv$ (mK km s$^{-1}$)\,$^a$} & Telescope & Reference \\
\hline
\endfirsthead
\caption{continued.}\\
\hline\hline
\multicolumn{1}{l}{Species} & \multicolumn{1}{c}{Transition} & \multicolumn{1}{c}{Frequency (MHz)} & \multicolumn{1}{c}{$\int T_A^* dv$ (mK km s$^{-1}$)\,$^a$} & Telescope & Reference \\
\hline
\endhead
\hline
\endfoot
\hline \hline
\multicolumn{6}{c}{TMC-1\,CP} \\
\hline \hline
C$_6$H & $^2\Pi_{3/2}$ $J=15/2-13/2$ $a$ &  20792.907 & 133(24)\,$^b$   & GBT        & \cite{Sakai2007} \\
       & $^2\Pi_{3/2}$ $J=15/2-13/2$ $b$ &  20794.475 & 112(22)\,$^b$   & GBT        & \cite{Sakai2007} \\
       & $^2\Pi_{3/2}$ $J=21/2-19/2$ $a$ &  29109.658 & 332.4(420)\,$^b$    & GBT        & \cite{Cordiner2013} \\
       & $^2\Pi_{3/2}$ $J=23/2-21/2$ $a$ &  31881.860 & 175.6(176)      & Yebes\,40m & This work \\
       & $^2\Pi_{3/2}$ $J=23/2-21/2$ $b$ &  31885.541 & 173.5(175)      & Yebes\,40m & This work \\
       & $^2\Pi_{3/2}$ $J=25/2-23/2$ $a$ &  34654.037 & 158.9(160)      & Yebes\,40m & This work \\
       & $^2\Pi_{3/2}$ $J=25/2-23/2$ $b$ &  34658.383 & 158.5(160)      & Yebes\,40m & This work \\
       & $^2\Pi_{3/2}$ $J=27/2-25/2$ $a$ &  37426.192 & 141.5(180)      & Yebes\,40m & This work \\
       & $^2\Pi_{3/2}$ $J=27/2-25/2$ $b$ &  37431.255 & 141.1(175)      & Yebes\,40m & This work \\
       & $^2\Pi_{3/2}$ $J=29/2-27/2$ $a$ &  40198.323 & 119.3(149)      & Yebes\,40m & This work \\
       & $^2\Pi_{3/2}$ $J=29/2-27/2$ $b$ &  40204.157 & 118.6(147)      & Yebes\,40m & This work \\
       & $^2\Pi_{3/2}$ $J=31/2-29/2$ $a$ &  42970.432 &  93.4(106)      & Yebes\,40m & This work \\
       & $^2\Pi_{3/2}$ $J=31/2-29/2$ $b$ &  42977.089 &  93.3(106)      & Yebes\,40m & This work \\
       & $^2\Pi_{3/2}$ $J=33/2-31/2$ $a$ &  45742.519 &  73.0(98)       & Yebes\,40m & This work \\
       & $^2\Pi_{3/2}$ $J=33/2-31/2$ $b$ &  45750.052 &  73.4(99)       & Yebes\,40m & This work \\
       & $^2\Pi_{3/2}$ $J=35/2-33/2$ $a$ &  48514.584 &  52.6(73)       & Yebes\,40m & This work \\
       & $^2\Pi_{3/2}$ $J=35/2-33/2$ $b$ &  48523.044 &  52.2(70)       & Yebes\,40m & This work \\
\hline
C$_4$H & $N=2-1$ $J=3/2-1/2$             &  19054.476 & 411.3(418)\,$^b$  & GBT        & \cite{Cordiner2013} \\
       & $N=4-3$ $J=9/2-7/2$             &  38049.654 & 1369(138)       & Yebes\,40m & This work \\
       & $N=4-3$ $J=7/2-5/2$             &  38088.461 & 1007(102)       & Yebes\,40m & This work \\
       & $N=5-4$ $J=11/2-9/2$            &  47566.792 & 1094(111)       & Yebes\,40m & This work \\
       & $N=5-4$ $J=9/2-7/2$             &  47605.496 &  864(87)        & Yebes\,40m & This work \\
       & $N=9-8$ $J=19/2-17/2$           &  85634.010 &  417(53)        & IRAM\,30m & \cite{Agundez2008} \\
       & $N=9-8$ $J=17/2-15/2$           &  85672.580 &  386(49)        & IRAM\,30m  & \cite{Agundez2008} \\
       & $N=10-9$ $J=21/2-19/2$          &  95150.393 &  251(26)        & IRAM\,30m  & This work \\
       & $N=10-9$ $J=19/2-17/2$          &  95188.947 &  243(26)        & IRAM\,30m  & This work \\
       & $N=11-10$ $J=23/2-21/2$         & 104666.568 &  111(12)        & IRAM\,30m  & This work \\
       & $N=11-10$ $J=21/2-19/2$         & 104705.108 &  105(13)        & IRAM\,30m  & This work \\
       & $N=12-11$ $J=25/2-23/2$         & 114182.523 &   60(8)         & IRAM\,30m  & This work \\
       & $N=12-11$ $J=23/2-21/2$         & 114221.023 &   47(6)         & IRAM\,30m  & This work \\
\hline
C$_8$H & $^2\Pi_{3/2}$ $J=53/2-51/2$ $a$ &  31093.035 &  6.0(7)         & Yebes\,40m & This work \\
       & $^2\Pi_{3/2}$ $J=53/2-51/2$ $b$ &  31093.415 &  4.4(6)         & Yebes\,40m & This work \\
       & $^2\Pi_{3/2}$ $J=55/2-53/2$ $a$ &  32266.325 &  4.3(6)         & Yebes\,40m & This work \\
       & $^2\Pi_{3/2}$ $J=55/2-53/2$ $b$ &  32266.735 &  4.2(6)         & Yebes\,40m & This work \\
       & $^2\Pi_{3/2}$ $J=57/2-55/2$ $a$ &  33439.612 &  3.5(5)         & Yebes\,40m & This work \\
       & $^2\Pi_{3/2}$ $J=57/2-55/2$ $b$ &  33440.052 &  3.4(6)         & Yebes\,40m & This work \\
       & $^2\Pi_{3/2}$ $J=59/2-57/2$ $b$ &  34613.367 &  2.7(3)         & Yebes\,40m & This work \\
       & $^2\Pi_{3/2}$ $J=61/2-59/2$ $a$ &  35786.176 &  3.0(4)         & Yebes\,40m & This work \\
       & $^2\Pi_{3/2}$ $J=61/2-59/2$ $b$ &  35786.679 &  2.4(3)         & Yebes\,40m & This work \\
       & $^2\Pi_{3/2}$ $J=63/2-61/2$ $a$ &  36959.452 &  2.3(3)         & Yebes\,40m & This work \\
       & $^2\Pi_{3/2}$ $J=63/2-61/2$ $b$ &  36959.989 &  2.2(3)         & Yebes\,40m & This work \\
       & $^2\Pi_{3/2}$ $J=65/2-63/2$ $a$ &  38132.725 &  1.7(2)         & Yebes\,40m & This work \\
       & $^2\Pi_{3/2}$ $J=65/2-63/2$ $b$ &  38133.297 &  1.5(2)         & Yebes\,40m & This work \\
       & $^2\Pi_{3/2}$ $J=67/2-65/2$ $a$ &  39305.995 &  1.4(2)         & Yebes\,40m & This work \\
       & $^2\Pi_{3/2}$ $J=67/2-65/2$ $b$ &  39306.602 &  1.4(2)         & Yebes\,40m & This work \\
       & $^2\Pi_{3/2}$ $J=69/2-67/2$ $a$ &  40479.260 &  1.2(2)         & Yebes\,40m & This work \\
       & $^2\Pi_{3/2}$ $J=69/2-67/2$ $b$ &  40479.904 &  1.2(2)         & Yebes\,40m & This work \\
       & $^2\Pi_{3/2}$ $J=71/2-69/2$ $a$ &  41652.522 &  0.8(1)         & Yebes\,40m & This work \\
       & $^2\Pi_{3/2}$ $J=71/2-69/2$ $b$ &  41653.203 &  0.9(1)         & Yebes\,40m & This work \\
       & $^2\Pi_{3/2}$ $J=73/2-71/2$ $a$ &  42825.779 &  0.7(1)         & Yebes\,40m & This work \\
       & $^2\Pi_{3/2}$ $J=73/2-71/2$ $b$ &  42826.499 &  0.7(1)         & Yebes\,40m & This work \\
\hline
C$_3$N & $N=4-3$ $J=9/2-7/2$             &  39571.347 & 332(34)         & Yebes\,40m & This work \\
       & $N=4-3$ $J=7/2-5/2$             &  39590.181 & 240(25)         & Yebes\,40m & This work \\
       & $N=5-4$ $J=11/2-9/2$            &  49466.421 & 244(25)         & Yebes\,40m & This work \\
       & $N=5-4$ $J=9/2-7/2$             &  49485.224 & 198(20)         & Yebes\,40m & This work \\
       & $N=9-8$ $J=19/2-17/2$           &  89045.583 &  64.2(73)       & IRAM\,30m  & This work \\
       & $N=9-8$ $J=17/2-15/2$           &  89064.347 &  58.6(68)       & IRAM\,30m  & This work \\
       & $N=10-9$ $J=21/2-19/2$          &  98940.087 &  28.1(36)       & IRAM\,30m  & This work \\
       & $N=10-9$ $J=19/2-17/2$          &  98958.770 &  22.7(30)       & IRAM\,30m  & This work \\
       & $N=11-10$ $J=23/2-21/2$         & 108834.254 &  11.6(24)       & IRAM\,30m  & This work \\
       & $N=11-10$ $J=21/2-19/2$         & 108853.012 &  21.2(35)       & IRAM\,30m  & This work \\
\hline
C$_5$N & $N=12-11$ $J=25/2-23/2$         &  33668.234 &  5.6(7)         & Yebes\,40m & This work \\
       & $N=12-11$ $J=23/2-21/2$         &  33678.966 &  5.9(7)         & Yebes\,40m & This work \\
       & $N=13-12$ $J=27/2-25/2$         &  36474.308 &  5.8(7)         & Yebes\,40m & This work \\
       & $N=13-12$ $J=25/2-23/2$         &  36485.042 &  5.5(7)         & Yebes\,40m & This work \\
       & $N=14-13$ $J=29/2-27/2$         &  39280.369 &  5.1(7)         & Yebes\,40m & This work \\
       & $N=14-13$ $J=27/2-25/2$         &  39291.105 &  5.0(7)         & Yebes\,40m & This work \\
       & $N=15-14$ $J=31/2-29/2$         &  42086.415 &  4.7(6)         & Yebes\,40m & This work \\
       & $N=15-14$ $J=29/2-27/2$         &  42097.151 &  4.4(6)         & Yebes\,40m & This work \\
       & $N=16-15$ $J=33/2-31/2$         &  44892.444 &  4.6(6)         & Yebes\,40m & This work \\
       & $N=16-15$ $J=31/2-29/2$         &  44903.182 &  4.4(6)         & Yebes\,40m & This work \\
       & $N=17-16$ $J=35/2-33/2$         &  47698.457 &  3.7(5)         & Yebes\,40m & This work \\
       & $N=17-16$ $J=33/2-31/2$         &  47709.196 &  3.4(5)         & Yebes\,40m & This work \\
\hline \hline
\multicolumn{6}{c}{Lupus-1A} \\
\hline \hline
C$_6$H & $^2\Pi_{3/2}$ $J=15/2-13/2$ $a$ &  20792.907 & 114(14)\,$^b$    & GBT        & \cite{Sakai2010} \\
       & $^2\Pi_{3/2}$ $J=15/2-13/2$ $b$ &  20794.475 & 131(16)\,$^b$    & GBT        & \cite{Sakai2010} \\
       & $^2\Pi_{3/2}$ $J=23/2-21/2$ $a$ &  31881.860 & 150.3(166)      & Yebes\,40m & This work \\
       & $^2\Pi_{3/2}$ $J=23/2-21/2$ $b$ &  31885.541 & 153.1(163)      & Yebes\,40m & This work \\
       & $^2\Pi_{3/2}$ $J=25/2-23/2$ $a$ &  34654.037 & 151.6(161)      & Yebes\,40m & This work \\
       & $^2\Pi_{3/2}$ $J=25/2-23/2$ $b$ &  34658.383 & 150.0(159)      & Yebes\,40m & This work \\
       & $^2\Pi_{3/2}$ $J=27/2-25/2$ $a$ &  37426.192 & 140.3(143)      & Yebes\,40m & This work \\
       & $^2\Pi_{3/2}$ $J=27/2-25/2$ $b$ &  37431.255 & 141.0(148)      & Yebes\,40m & This work \\
       & $^2\Pi_{3/2}$ $J=29/2-27/2$ $a$ &  40198.323 & 126.2(134)      & Yebes\,40m & This work \\
       & $^2\Pi_{3/2}$ $J=29/2-27/2$ $b$ &  40204.157 & 124.8(130)      & Yebes\,40m & This work \\
       & $^2\Pi_{3/2}$ $J=31/2-29/2$ $a$ &  42970.432 & 115.5(123)      & Yebes\,40m & This work \\
       & $^2\Pi_{3/2}$ $J=31/2-29/2$ $b$ &  42977.089 & 114.9(123)      & Yebes\,40m & This work \\
       & $^2\Pi_{3/2}$ $J=33/2-31/2$ $a$ &  45742.519 &  90.7(125)      & Yebes\,40m & This work \\
       & $^2\Pi_{3/2}$ $J=33/2-31/2$ $b$ &  45750.052 &  91.3(128)      & Yebes\,40m & This work \\
       & $^2\Pi_{3/2}$ $J=35/2-33/2$ $a$ &  48514.584 &  73.6(109)      & Yebes\,40m & This work \\
       & $^2\Pi_{3/2}$ $J=35/2-33/2$ $b$ &  48523.044 &  66.9(103)      & Yebes\,40m & This work \\
\hline
C$_4$H & $N=4-3$ $J=9/2-7/2$             &  38049.654 & 1219(123)       & Yebes\,40m & This work \\
       & $N=4-3$ $J=7/2-5/2$             &  38088.461 &  921(94)        & Yebes\,40m & This work \\
       & $N=5-4$ $J=11/2-9/2$            &  47566.792 & 1123(114)       & Yebes\,40m & This work \\
       & $N=5-4$ $J=9/2-7/2$             &  47605.496 &  846(86)        & Yebes\,40m & This work \\
       & $N=8-7$ $J=17/2-15/2$           &  76117.439 & 1124(114)       & IRAM\,30m  & This work \\
       & $N=8-7$ $J=15/2-13/2$           &  76156.028 & 1024(104)       & IRAM\,30m  & This work \\
       & $N=9-8$ $J=19/2-17/2$           &  85634.010 &  779(83)        & IRAM\,30m  & This work \\
       & $N=9-8$ $J=17/2-15/2$           &  85672.580 &  730(77)        & IRAM\,30m  & This work \\
       & $N=11-10$ $J=23/2-21/2$         & 104666.568 &  349(39)        & IRAM\,30m  & This work \\
       & $N=11-10$ $J=21/2-19/2$         & 104705.108 &  334(38)        & IRAM\,30m  & This work \\
\hline
C$_8$H & $^2\Pi_{3/2}$ $J=33/2-31/2$ $a$ &  19359.975 & 10(2)\,$^b$     & GBT        & \cite{Sakai2010} \\
       & $^2\Pi_{3/2}$ $J=33/2-31/2$ $b$ &  19360.123 & 9(2)\,$^b$      & GBT        & \cite{Sakai2010} \\
\hline
C$_3$N & $N=4-3$ $J=9/2-7/2$             &  39571.347 & 251(30)         & Yebes\,40m & This work \\
       & $N=4-3$ $J=7/2-5/2$             &  39590.181 & 175(19)         & Yebes\,40m & This work \\
       & $N=5-4$ $J=11/2-9/2$            &  49466.421 & 177(19)         & Yebes\,40m & This work \\
       & $N=5-4$ $J=9/2-7/2$             &  49485.224 & 138(15)         & Yebes\,40m & This work \\
       & $N=9-8$ $J=19/2-17/2$           &  89045.583 & 141.5(150)      & IRAM\,30m  & This work \\
       & $N=9-8$ $J=17/2-15/2$           &  89064.347 & 126.7(136)      & IRAM\,30m  & This work \\
       & $N=10-9$ $J=21/2-19/2$          &  98940.087 &  74.6(83)       & IRAM\,30m  & This work \\
       & $N=10-9$ $J=19/2-17/2$          &  98958.770 &  66.0(74)       & IRAM\,30m  & This work \\
\hline
C$_5$N & $N=12-11$ $J=25/2-23/2$         &  33668.234 &  4.5(12)        & Yebes\,40m & This work \\
       & $N=12-11$ $J=23/2-21/2$         &  33678.966 &  7.0(14)        & Yebes\,40m & This work \\
       & $N=13-12$ $J=27/2-25/2$         &  36474.308 &  4.8(11)        & Yebes\,40m & This work \\
       & $N=13-12$ $J=25/2-23/2$         &  36485.042 &  5.7(11)        & Yebes\,40m & This work \\
       & $N=14-13$ $J=29/2-27/2$         &  39280.369 &  7.8(24)        & Yebes\,40m & This work \\ 
       & $N=14-13$ $J=27/2-25/2$         &  39291.105 &  5.7(15)        & Yebes\,40m & This work \\ 
       & $N=15-14$ $J=31/2-29/2$         &  42086.415 &  4.1(9)         & Yebes\,40m & This work \\
       & $N=15-14$ $J=29/2-27/2$         &  42097.151 &  4.8(11)        & Yebes\,40m & This work \\
       & $N=16-15$ $J=33/2-31/2$         &  44892.444 &  3.2(9)         & Yebes\,40m & This work \\
       & $N=16-15$ $J=31/2-29/2$         &  44903.182 &  1.8(8)\,$^d$         & Yebes\,40m & This work \\ 
\hline \hline
\multicolumn{6}{c}{L1527} \\
\hline \hline
C$_6$H & $^2\Pi_{3/2}$ $J=15/2-13/2$ $a$ &  20792.907 & 24(5)\,$^b$     & GBT        & \cite{Sakai2007} \\
       & $^2\Pi_{3/2}$ $J=15/2-13/2$ $b$ &  20794.475 & 21(5)\,$^b$     & GBT        & \cite{Sakai2007} \\
       & $^2\Pi_{3/2}$ $J=23/2-21/2$ $a$ &  31881.860 &  34.8(75)       & Yebes\,40m & This work \\
       & $^2\Pi_{3/2}$ $J=23/2-21/2$ $b$ &  31885.541 &  26.0(59)       & Yebes\,40m & This work \\
       & $^2\Pi_{3/2}$ $J=25/2-23/2$ $a$ &  34654.037 &  29.3(34)       & Yebes\,40m & This work \\
       & $^2\Pi_{3/2}$ $J=25/2-23/2$ $b$ &  34658.383 &  31.8(37)       & Yebes\,40m & This work \\
       & $^2\Pi_{3/2}$ $J=27/2-25/2$ $a$ &  37426.192 &  31.7(46)       & Yebes\,40m & This work \\
       & $^2\Pi_{3/2}$ $J=27/2-25/2$ $b$ &  37431.255 &  32.2(51)       & Yebes\,40m & This work \\
       & $^2\Pi_{3/2}$ $J=29/2-27/2$ $a$ &  40198.323 &  32.7(50)       & Yebes\,40m & This work \\
       & $^2\Pi_{3/2}$ $J=29/2-27/2$ $b$ &  40204.157 &  32.3(48)       & Yebes\,40m & This work \\
       & $^2\Pi_{3/2}$ $J=31/2-29/2$ $a$ &  42970.432 &  30.2(47)       & Yebes\,40m & This work \\
       & $^2\Pi_{3/2}$ $J=31/2-29/2$ $b$ &  42977.089 &  31.1(49)       & Yebes\,40m & This work \\
       & $^2\Pi_{3/2}$ $J=33/2-31/2$ $a$ &  45742.519 &  30.5(48)       & Yebes\,40m & This work \\
       & $^2\Pi_{3/2}$ $J=33/2-31/2$ $b$ &  45750.052 &  31.3(49)       & Yebes\,40m & This work \\
       & $^2\Pi_{3/2}$ $J=35/2-33/2$ $a$ &  48514.584 &  27.3(48)       & Yebes\,40m & This work \\
       & $^2\Pi_{3/2}$ $J=35/2-33/2$ $b$ &  48523.044 &  26.9(47)       & Yebes\,40m & This work \\
\hline
C$_4$H & $N=4-3$ $J=9/2-7/2$             &  38049.654 & 388(39)         & Yebes\,40m & This work \\
       & $N=4-3$ $J=7/2-5/2$             &  38088.461 & 295(30)         & Yebes\,40m & This work \\
       & $N=5-4$ $J=11/2-9/2$            &  47566.792 & 434(44)         & Yebes\,40m & This work \\
       & $N=5-4$ $J=9/2-7/2$             &  47605.496 & 347(35)         & Yebes\,40m & This work \\
       & $N=9-8$ $J=19/2-17/2$           &  85634.010 & 747(86)         & IRAM\,30m  & \cite{Agundez2008} \\ 
       & $N=9-8$ $J=17/2-15/2$           &  85672.580 & 712(82)         & IRAM\,30m  & \cite{Agundez2008} \\
       & $N=11-10$ $J=23/2-21/2$         & 104666.568 & 542(64)         & IRAM\,30m  & \cite{Agundez2008} \\
       & $N=11-10$ $J=21/2-19/2$         & 104705.108 & 487(59)         & IRAM\,30m  & \cite{Agundez2008} \\
       & $N=12-11$ $J=25/2-23/2$         & 114182.523 & 462(59)         & IRAM\,30m  & \cite{Agundez2008} \\
       & $N=12-11$ $J=23/2-21/2$         & 114221.023 & 406(53)         & IRAM\,30m  & \cite{Agundez2008} \\
\hline \hline
\multicolumn{6}{c}{L483} \\
\hline \hline
C$_6$H & $^2\Pi_{3/2}$ $J=23/2-21/2$ $a$ &  31881.860 &  29.4(34)       & Yebes\,40m & This work \\
       & $^2\Pi_{3/2}$ $J=23/2-21/2$ $b$ &  31885.541 &  31.0(36)       & Yebes\,40m & This work \\
       & $^2\Pi_{3/2}$ $J=25/2-23/2$ $a$ &  34654.037 &  28.4(32)       & Yebes\,40m & This work \\
       & $^2\Pi_{3/2}$ $J=25/2-23/2$ $b$ &  34658.383 &  27.7(31)       & Yebes\,40m & This work \\
       & $^2\Pi_{3/2}$ $J=27/2-25/2$ $a$ &  37426.192 &  26.2(29)       & Yebes\,40m & This work \\
       & $^2\Pi_{3/2}$ $J=27/2-25/2$ $b$ &  37431.255 &  26.2(30)       & Yebes\,40m & This work \\
       & $^2\Pi_{3/2}$ $J=29/2-27/2$ $a$ &  40198.323 &  24.4(28)       & Yebes\,40m & This work \\
       & $^2\Pi_{3/2}$ $J=29/2-27/2$ $b$ &  40204.157 &  23.2(27)       & Yebes\,40m & This work \\
       & $^2\Pi_{3/2}$ $J=31/2-29/2$ $a$ &  42970.432 &  19.7(23)       & Yebes\,40m & This work \\
       & $^2\Pi_{3/2}$ $J=31/2-29/2$ $b$ &  42977.089 &  20.4(24)       & Yebes\,40m & This work \\
       & $^2\Pi_{3/2}$ $J=33/2-31/2$ $a$ &  45742.519 &  13.6(22)       & Yebes\,40m & This work \\
       & $^2\Pi_{3/2}$ $J=33/2-31/2$ $b$ &  45750.052 &  14.2(21)       & Yebes\,40m & This work \\
       & $^2\Pi_{3/2}$ $J=35/2-33/2$ $a$ &  48514.584 &  13.0(23)       & Yebes\,40m & This work \\
       & $^2\Pi_{3/2}$ $J=35/2-33/2$ $b$ &  48523.044 &  13.7(24)       & Yebes\,40m & This work \\
\hline
C$_4$H & $N=4-3$ $J=9/2-7/2$             &  38049.654 & 470(48)         & Yebes\,40m & This work \\
       & $N=4-3$ $J=7/2-5/2$             &  38088.461 & 356(36)         & Yebes\,40m & This work \\
       & $N=5-4$ $J=11/2-9/2$            &  47566.792 & 439(50)         & Yebes\,40m & This work \\
       & $N=5-4$ $J=9/2-7/2$             &  47605.496 & 352(36)         & Yebes\,40m & This work \\
       & $N=8-7$ $J=17/2-15/2$           &  76117.439 & 375(38)        & IRAM\,30m  & This work \\
       & $N=8-7$ $J=15/2-13/2$           &  76156.028 & 337(35)       & IRAM\,30m  & This work \\
       & $N=9-8$ $J=19/2-17/2$           &  85634.010 & 272(27)         & IRAM\,30m  & \cite{Agundez2019} \\
       & $N=9-8$ $J=17/2-15/2$           &  85672.580 & 249(24)         & IRAM\,30m  & \cite{Agundez2019} \\
       & $N=10-9$ $J=21/2-19/2$          &  95150.393 & 157(15)         & IRAM\,30m  & \cite{Agundez2019} \\
       & $N=10-9$ $J=19/2-17/2$          &  95188.947 & 147(14)         & IRAM\,30m  & \cite{Agundez2019} \\
       & $N=11-10$ $J=23/2-21/2$         & 104666.568 & 110(10)         & IRAM\,30m  & \cite{Agundez2019} \\
       & $N=11-10$ $J=21/2-19/2$         & 104705.108 & 100(9)          & IRAM\,30m  & \cite{Agundez2019} \\
       & $N=12-11$ $J=25/2-23/2$         & 114182.523 &  64(6)          & IRAM\,30m  & \cite{Agundez2019} \\
       & $N=12-11$ $J=23/2-21/2$         & 114221.023 &  64(6)          & IRAM\,30m  & \cite{Agundez2019} \\
\hline \hline
\multicolumn{6}{c}{L1495B} \\
\hline \hline
C$_6$H & $^2\Pi_{3/2}$ $J=13/2-11/2$ $a$ &  18020.606 & 55(10)\,$^c$           & GBT        & \cite{Gupta2009} \\
       & $^2\Pi_{3/2}$ $J=13/2-11/2$ $b$ &  18021.783 & 55(10)\,$^c$           & GBT        & \cite{Gupta2009} \\
       & $^2\Pi_{3/2}$ $J=21/2-19/2$ $a$ &  29109.658 & 141.6(164)\,$^b$ & GBT        & \cite{Cordiner2013} \\
       & $^2\Pi_{3/2}$ $J=23/2-21/2$ $a$ &  31881.860 &  51.9(59)       & Yebes\,40m & This work \\
       & $^2\Pi_{3/2}$ $J=23/2-21/2$ $b$ &  31885.541 &  47.9(53)       & Yebes\,40m & This work \\
       & $^2\Pi_{3/2}$ $J=25/2-23/2$ $a$ &  34654.037 &  46.8(52)       & Yebes\,40m & This work \\
       & $^2\Pi_{3/2}$ $J=27/2-25/2$ $a$ &  37426.192 &  45.4(51)       & Yebes\,40m & This work \\
       & $^2\Pi_{3/2}$ $J=27/2-25/2$ $b$ &  37431.255 &  42.8(49)       & Yebes\,40m & This work \\
       & $^2\Pi_{3/2}$ $J=29/2-27/2$ $a$ &  40198.323 &  36.2(42)       & Yebes\,40m & This work \\
       & $^2\Pi_{3/2}$ $J=29/2-27/2$ $b$ &  40204.157 &  37.7(42)       & Yebes\,40m & This work \\
       & $^2\Pi_{3/2}$ $J=31/2-29/2$ $a$ &  42970.432 &  33.3(40)       & Yebes\,40m & This work \\
       & $^2\Pi_{3/2}$ $J=31/2-29/2$ $b$ &  42977.089 &  33.6(40)       & Yebes\,40m & This work \\
       & $^2\Pi_{3/2}$ $J=33/2-31/2$ $a$ &  45742.519 &  24.7(38)       & Yebes\,40m & This work \\
       & $^2\Pi_{3/2}$ $J=33/2-31/2$ $b$ &  45750.052 &  24.0(35)       & Yebes\,40m & This work \\
       & $^2\Pi_{3/2}$ $J=35/2-33/2$ $a$ &  48514.584 &  19.4(32)       & Yebes\,40m & This work \\
       & $^2\Pi_{3/2}$ $J=35/2-33/2$ $b$ &  48523.044 &  18.6(33)       & Yebes\,40m & This work \\
\hline \hline
\multicolumn{6}{c}{L1544} \\
\hline \hline
C$_6$H & $^2\Pi_{3/2}$ $J=13/2-11/2$ $a$ &  18020.606 & 51(11)           & GBT        & \cite{Gupta2009} \\
       & $^2\Pi_{3/2}$ $J=13/2-11/2$ $b$ &  18021.783 & 50(11)           & GBT        & \cite{Gupta2009} \\
       & $^2\Pi_{3/2}$ $J=23/2-21/2$ $a$ &  31881.860 &  23.8(36)       & Yebes\,40m & This work \\
       & $^2\Pi_{3/2}$ $J=23/2-21/2$ $b$ &  31885.541 &  30.0(44)       & Yebes\,40m & This work \\
       & $^2\Pi_{3/2}$ $J=25/2-23/2$ $a$ &  34654.037 &  25.7(39)       & Yebes\,40m & This work \\
       & $^2\Pi_{3/2}$ $J=25/2-23/2$ $b$ &  34658.383 &  31.6(48)       & Yebes\,40m & This work \\
       & $^2\Pi_{3/2}$ $J=27/2-25/2$ $a$ &  37426.192 &  23.3(36)       & Yebes\,40m & This work \\
       & $^2\Pi_{3/2}$ $J=27/2-25/2$ $b$ &  37431.255 &  19.9(34)       & Yebes\,40m & This work \\
       & $^2\Pi_{3/2}$ $J=29/2-27/2$ $b$ &  40204.157 &  18.0(31)       & Yebes\,40m & This work \\
       & $^2\Pi_{3/2}$ $J=31/2-29/2$ $a$ &  42970.432 &  13.6(26)       & Yebes\,40m & This work \\
       & $^2\Pi_{3/2}$ $J=31/2-29/2$ $b$ &  42977.089 &  12.1(23)       & Yebes\,40m & This work \\
\hline \hline
\multicolumn{6}{c}{L1521F} \\
\hline \hline
C$_6$H & $^2\Pi_{3/2}$ $J=13/2-11/2$ $a$ &  18020.606 & 36(10)           & GBT        & \cite{Gupta2009} \\
       & $^2\Pi_{3/2}$ $J=13/2-11/2$ $b$ &  18021.783 & 26(9)           & GBT        & \cite{Gupta2009} \\
\hline \hline
\multicolumn{6}{c}{L1251A} \\
\hline \hline
C$_6$H & $^2\Pi_{3/2}$ $J=21/2-19/2$ $a$ &  29109.658 & 36(8)           & GBT        & \cite{Cordiner2011} \\
       & $^2\Pi_{3/2}$ $J=21/2-19/2$ $b$ &  29112.730 & 35(8)           & GBT        & \cite{Cordiner2011} \\
       & $^2\Pi_{3/2}$ $J=21/2-19/2$ $a$ &  29109.658 & 43.6(65)\,$^b$  & GBT        & \cite{Cordiner2013} \\
\hline \hline
\multicolumn{6}{c}{L1512} \\
\hline \hline
C$_6$H & $^2\Pi_{3/2}$ $J=13/2-11/2$ $a$ &  18020.606 & 20(7)\,$^c$           & GBT        & \cite{Gupta2009} \\
       & $^2\Pi_{3/2}$ $J=13/2-11/2$ $b$ &  18021.783 & 20(7)\,$^c$           & GBT        & \cite{Gupta2009} \\
       & $^2\Pi_{3/2}$ $J=21/2-19/2$ $a$ &  29109.658 & 27(5)           & GBT        & \cite{Cordiner2011} \\
       & $^2\Pi_{3/2}$ $J=21/2-19/2$ $b$ &  29112.730 & 28(5)           & GBT        & \cite{Cordiner2011} \\
       & $^2\Pi_{3/2}$ $J=21/2-19/2$ $a$ &  29109.658 & 26.3(35)\,$^b$   & GBT        & \cite{Cordiner2013} \\
\hline \hline
\multicolumn{6}{c}{L1172} \\
\hline \hline
C$_6$H & $^2\Pi_{3/2}$ $J=21/2-19/2$ $a$ &  29109.658 & 41.1(57)\,$^b$  & GBT        & \cite{Cordiner2013} \\
\hline \hline
\multicolumn{6}{c}{L1389} \\
\hline \hline
C$_6$H & $^2\Pi_{3/2}$ $J=13/2-11/2$ $a$ &  18020.606 & 10(6)\,$^c$           & GBT        & \cite{Gupta2009} \\
       & $^2\Pi_{3/2}$ $J=13/2-11/2$ $b$ &  18021.783 & 10(6)\,$^c$           & GBT        & \cite{Gupta2009} \\
       & $^2\Pi_{3/2}$ $J=21/2-19/2$ $a$ &  29109.658 & 27.1(40)\,$^b$  & GBT        & \cite{Cordiner2013} \\
\hline \hline
\multicolumn{6}{c}{TMC-1\,C} \\
\hline \hline
C$_6$H & $^2\Pi_{3/2}$ $J=21/2-19/2$ $a$ &  29109.658 & 88.1(105)\,$^b$  & GBT        & \cite{Cordiner2013} \\
\end{longtable}
\tablenotea{\\
$^a$\,Unless otherwise stated, the intensity scale is antenna temperature ($T_A^*$). It can be converted to main beam brightness temperature ($T_{\rm mb}$) by dividing by $B_{\rm eff}$/$F_{\rm eff}$ (see caption of Table~\ref{table:lines_anion}. The error in $\int T_A^* dv$ includes the contributions from the Gaussian fit and from calibration (assumed to be 10\,\%). $^b$\,Intensity scale is $T_{\rm mb}$. $^c$\,Intensity distributed equally among the two fine components. $^d$\,Marginal detection.
}

\end{document}